\documentclass[
aps,
reprint,
prb,
superscriptaddress,
amsmath,
amssymb,
floatfix,
longbibliography
]{revtex4-2}
\usepackage{bm}
\usepackage{graphicx}
\usepackage{latexsym}
\usepackage{array}
\usepackage{amsfonts}
\usepackage[free-standing-units]{siunitx}

\usepackage{mathtools}  
\usepackage{makecell}   
\usepackage{hhline}     
\usepackage{amsthm}
\usepackage{amscd}
\usepackage{xcolor}
\renewcommand\vec{\boldsymbol}

\definecolor{orange}{rgb}{1,0.5,0}
\definecolor{goodgreen}{rgb}{0.1,0.5,0}
\definecolor{goodred}{rgb}{0.7,0,0}
\usepackage[colorlinks,urlcolor=goodgreen,citecolor=blue,linkcolor=goodred]{hyperref}

\newcommand{\orcid}[1]{\href{https://orcid.org/#1}{\includegraphics[width=8pt]{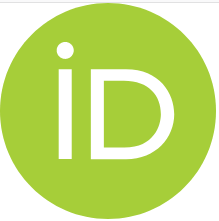}}}

\usepackage{verbatim}
\usepackage{hyperref}
\usepackage[normalem]{ulem}
\graphicspath{{images/}}

\usepackage{etoolbox}
\apptocmd{\sloppy}{\hbadness 10000\relax}{}{}

\begin{document}

\title{Interfacial spin-orbit coupling in superconducting hybrid systems}

\author{A. A. Mazanik\orcid{0000-0001-6389-8653}}
\email{andrei.mazanik@csic.es}
\affiliation{Centro de Física de Materiales (CFM-MPC) Centro Mixto CSIC-UPV/EHU,
E-20018 Donostia-San Sebastián, Spain}

\author{Tim Kokkeler\orcid{0000-0001-8681-3376}}
\email{tim.kokkeler@dipc.org}

\affiliation{Donostia International Physics Center (DIPC), 20018 Donostia--San Sebastián, Spain}
\affiliation{University of Twente, 7522 NB Enschede, The Netherlands}

\author{I. V. Tokatly\orcid{0000-0001-6288-0689}}
\email{ilya.tokatly@ehu.es}

\affiliation{Donostia International Physics Center (DIPC), 20018 Donostia--San
Sebastián, Spain}
\affiliation{Departamento de Polimeros y Materiales
Avanzados, Universidad del Pais Vasco UPV/EHU, 20018 Donostia-San Sebastian, Spain}
\affiliation{IKERBASQUE, Basque Foundation for Science, 48009 Bilbao, Spain}

\author{F. Sebastian Bergeret\orcid{0000-0001-6007-4878}}
\email{fs.bergeret@csic.es}

\affiliation{Centro de Física de Materiales (CFM-MPC) Centro Mixto CSIC-UPV/EHU,
E-20018 Donostia-San Sebastián, Spain}
\affiliation{Donostia International Physics Center (DIPC), 20018 Donostia--San
Sebastián, Spain}

\date{\today}
\begin{abstract}

We investigate the effects of interfacial spin-orbit coupling (ISOC) on superconductors, focusing on its impact on electronic transport and spin-charge conversion. Using a symmetry-based nonlinear sigma model, we derive effective boundary conditions for the Usadel and Maxwell equations that account for the spin-galvanic effect, spin relaxation, and spin precession. This approach allows for the analysis of various interfaces without relying on specific microscopic models. We apply these boundary conditions to derive ISOC-induced terms  in the Ginzburg-Landau functional, which is then used to compute the critical temperature of superconducting films with ISOC subjected to an external magnetic field. 
Our findings show that, contrary to a recent prediction, the critical temperature of a film cannot be enhanced by an external magnetic field.
Additionally, we demonstrate that the combination of ISOC and an external magnetic field leads to a superconducting diode effect.   Its efficiency strongly depends on the interplay between the spin-galvanic and the spin relaxation terms. Our results provide a framework for understanding ISOC in superconducting systems and highlight the potential for optimizing diode efficiency through careful interface engineering.

\end{abstract}

\maketitle


\section{Introduction}
Charge-spin interconversion in materials with spin-orbit coupling (SOC) has been a prominent research topic over the past decade \cite{sinova2015spin,valenzuela2006direct,ganichev2002spin}. Magnetoelectric effects, such as spin Hall and spin-galvanic effects, manifest not only in normal conductors but also in superconducting materials \cite{buzdin2008direct,konschelle2015theory,bergeret2016manifestation,huang2018extrinsic,he2020magnetoelectric,virtanen2021magnetoelectric,virtanen2022nonlinear,mel2022superconducting,Shukrinov_2022,Bobkova_2022}. These effects are closely related to nonreciprocal electronic transport in superconductors, with the superconducting diode effect emerging as one of the most promising phenomena \cite{nadeem2023superconducting,daido2022intrinsic,ando2020observation,he2022phenomenological,ilic2022theory,kokkeler2024nonreciprocal,ilic2024superconducting,pal2022josephson,wu2022field,kindiak2024reduced,costa2023sign,costa2023sign}. 

SOC may arise in bulk systems, but is particularly ubiquitous near boundaries, because of the breaking of inversion symmetry. Charge-spin interconnection due to interface spin-orbit coupling (ISOC) has been observed in several experimental works \cite{rojas2016spin,kondou2016fermi,lesne2016highly,vaz2019mapping,sanz2020quantification,pham2021large,Groen23}. 
In some theoretical works, ISOC is incorporated in the normal state by assuming a two-dimensional electron gas with Rashba SOC \cite{rashba1960properties} that exists between the two materials \cite{kim2013chirality,koo2020rashba,sanchez2013spin,bihlmayer2022rashba,nomura2015temperature,chen2015spin,borge2014spin,noh2015controlling, Amin20SOT}.

However, while such an approach can be well motivated in ideal cases, real interfaces between two materials are not perfect, and it is not clear which features of such two-dimensional electron gas are robust against perturbations and which are not.
Alternatively, the description of mesoscopic hybrid structures with disorder can be achieved by combining drift-diffusion theory with effective boundary conditions \cite{borge2017ballistic,borge2019boundary}. These boundary conditions incorporate both spin-charge interconversion and spin relaxation. This approach has been successfully used to describe experiments on ISOC-induced magnetoelectric effects in normal systems \cite{sanz2020quantification,pham2021large,Groen23}.

In the superconducting state, the effect of ISOC on electronic transport has not been studied as extensively as in the normal state.  For specific models of interfaces with spin-orbit coupling, the resulting boundary conditions were derived in \cite{Linder19boundary, Linder22boundary}. In these works, the spin-orbit coupling was introduced respectively as a 2D Rashba layer \cite{Linder19boundary} and the inclusion of a 2D covariant derivative at the interface. In both cases, terms were kept to the second order in ISOC strength, and it was found that if one of the materials is an insulator, the only effect of ISOC is relaxation. This is in line with similar models for linear bulk spin-orbit coupling \cite{virtanen2022nonlinear}, in which it was found that for linear spin-charge coupling, spin-charge conversion through the spin-galvanic effect appears only to third order in the spin-charge coupling strength. 

To avoid dependence on a specific model of the interface and to capture all effects due to spin-charge coupling, phenomenological theories are best suited. Within superconductivity, the most well-known phenomenological theory is the Ginzburg-Landau functional. The Ginzburg-Landau functional in the presence of ISOC was used in different works; see \cite{melnikov2022superconducting} and references therein.  For instance, the Ginzburg-Landau functional with an ISOC-induced term was recently utilized in \cite{devizorova2024interfacial} to predict an enhancement of the critical temperature upon the application of a magnetic field, which is in sharp contrast with the magnetic field dependence of the critical temperature in junctions without ISOC. The advantage of this type of phenomenological approach is its independence from any specific microscopic model of the interface, which makes it broadly applicable. 

However, the G-L approach comes with a few disadvantages. First of all, it can only describe superconductors close to their critical temperature and is therefore not suitable for the description at low temperatures, at which many experiments are carried out. Next to this, by going to the critical temperature, the superconducting coherence length diverges, and therefore the G-L theory should be treated with caution in systems of finite extent.
The phenomenological description of bulk spin-orbit coupling in the low temperature regime and on smaller scales was recently achieved by applying a phenomenological approach to the non-linear sigma model (NLSM) \cite{kokkeler2024universal}. Because of the NLSM structure, such models are valid for all temperatures, while they are spatially accurate on scales much larger than the scattering length, which is a considerably weaker condition than for G-L theory. Meanwhile, by using phenomenological arguments for spin-orbit coupling, all possible types of spin-orbit-induced terms for a given material were identified, thereby automatically capturing the spin-charge conversion induced by spin-orbit coupling.

In this work, we use the method developed in \cite{kokkeler2024universal} to derive the NLSM and describe normal metals and superconductors with ISOC at their boundaries. The latter is represented by constructing an effective boundary action that contains all symmetry-allowed effects, specifically spin precession, spin-galvanic effect, and spin relaxation. Using the saddle-point approximation of the derived NLSM, we find the bulk Usadel equation, and the boundary condition at the interface with an insulator with ISOC. With this, we provide the first description of low-temperature transport and spin-charge conversion in superconducting systems with ISOC.

Based on these equations we calculate the influence of ISOC and spin relaxation on the critical temperature and diode efficiency of superconductor/insulator junctions. We show that the critical temperature in such systems cannot increase with the application of a magnetic field, in contrast to the result of~\cite{devizorova2024interfacial}. 
To clarify this discrepancy, we derive the corresponding G-L functional and demonstrate that, to the second order, the spin-galvanic tensor increases the superconducting stiffness, that is, it becomes more costly to create phase gradients. This increase compensates any possible enhancement of the critical temperature in a magnetic field by the ISOC-induced Lifshitz invariant, thereby nullifying the effect. With decreasing thickness, the magnetic field dependence becomes weaker but, unlike in junctions without ISOC, the magnetic field dependence does not disappear, but instead converges to an ISOC strength dependent value. Thus, ISOC in fact makes thin materials more fragile against magnetic fields.

A consequence of the Lifshitz invariant that is not nullified, but only reduced by this increased stiffness, is the appearance of an equilibrium phase gradient for nonzero magnetic fields, which is usually indicative of a superconducting diode effects.
To explore this effect, we solve the Usadel equation for low temperatures for thin superconductors and show that the superconducting diode effect indeed exists.
We find that the strength of the superconducting diode effect crucially hinges on the relative strength of the spin-galvanic and spin relaxation terms, thereby emphasizing the importance of developing junctions with small relaxation. 
In particular, its efficiency may approach $\approx 20\%$ in thin films for reasonable intensity of the interface SOC in an in-plane external field of $ \approx 0.7 H_{c2}$ for low enough values of spin relaxation.  Here, $H_{c2}$ is the second critical field of the superconductor.

The structure of the paper is as follows. In the next section, Sec. \ref{sec:main results}, we summarize our main results.   
In Sec. \ref{sec:sigma_model}, we present the NLSM for systems with ISOC and derive the corresponding Usadel and Maxwell equations, as well as boundary conditions,  through the minimization of the action. These equations, one of the central results of this work, can be used to describe generic systems containing interfacial spin-orbit coupling.  In Sec. \ref{sec:GL}, we use these equations to consider the behavior of a thin superconductor with ISOC near $T_{c}$. We derive the G-L functional from the NLSM, compute the critical temperature, and show that, in the thin junction limit, the superconducting stiffness is altered by the ISOC. Lastly, in Sec. \ref{sec:Diode}, we consider the non-linear Usadel equation to calculate the diode efficiency of the system. We present our conclusions in the final section. 
Throughout the paper, we use units with $\hbar = k_{B} = 1$.


\section{Main Results}
\label{sec:main results}
Before delving into the detailed derivation and examples, we first summarize our main results.

Our goal is to describe the spectral and transport properties of normal and superconducting metals where spin-orbit coupling occurs solely at the interface with another material. In other words, transport within the metal is governed by the well-established Usadel equation for the Keldysh quasiclassical Green's function $\check g$:
\begin{align} \label{eq:Usadel_main}
    D\tilde{\nabla}(\check{g}\tilde{\nabla}\check{g}) = \left[\hat{\omega}_{t,t'}\check{\tau}_{3}+i\vec{h}\cdot\vec{\sigma}\tau_3+\Delta\check{\tau}_{2}e^{i\varphi\check{\tau}_{3}},\ \check{g} \right]\;,
\end{align}
where $D$ is the diffusion constant, $\vec{\sigma} = (\sigma_{x},\sigma_{y},\sigma_{z})$ is the vector of Pauli matrices in spin space, $\hat{\omega}_{t,t'} = \delta(t-t')\partial_{t}$ is the time-derivative, $\vec{h}$ is the exchange field in the material, $\Delta$ is the magnitude of the pair potential and $\varphi$ is its phase. 

Our goal is to derive an effective boundary condition for Eq.~(\ref{eq:Usadel_main}) that accounts for interfacial spin-orbit coupling effects, including charge-spin conversion and spin relaxation. A similar boundary condition was obtained for the normal state in Ref.~\cite{borge2019boundary} by incorporating all symmetry allowed terms associated with ISOC. Generalizing this boundary condition to the superconducting state is not straightforward and constitutes the central result of this work, with the derivation presented in the next section. 

The derived boundary condition at an isotropic  interface between the superconductor and a heavy atom insulator reads:
\begin{equation}\label{eq:Boundary_ISOC_main}
    \left.-Dn_k \check{g}\tilde{\nabla}_{k} \check{g} \right\vert_{z = 0} =  
 -\vec{\tilde{\nabla}}\cdot \vec{\mathcal{\check{J}}}^{S}+\mathcal{\check{T}}^{S}-2\Gamma_{ab}\left[\sigma_{a}\check{g}\sigma_{b},\ \check{g}\right]\;.   
\end{equation}
Here $n_k$ is the $k$ component of the vector normal to the interface. Sum over repeated indices is implied. 
The term $\Gamma_{ab}$ is the  spin-relaxation tensor. For an isotropic interface  
there is  only allowed  gyrotropic tensor allowed \cite{kokkeler2024nonreciprocal}: $n_{r}\varepsilon_{lkr}$, where $n_{r}$ is the normal to the interface and $\varepsilon_{lkr}$ is the fully antisymmetric third rank tensor.  This tensor enters  the expressions  for  the matrix surface current    
 $\vec{\mathcal{\check{J}}}^S$, 
and for the matrix surface torque $\mathcal{\check{T}}^S$, as 
\begin{align}
     \label{eq:matrixJ_Main_Results}
 \mathcal{\check{J}}^{S}_{k} &= \frac{i}{2} \alpha n_r \epsilon_{kjr} \left[\check{g}\sigma_{j},\ \check{g}\right] + \nonumber\\&+\frac{i}{16}\gamma n_{r}\epsilon_{ijl}\epsilon_{lkr}\left\{\left[\sigma_{i},\ \check{g}\right],\ \sigma_{j}+\check{g}\sigma_{j}\check{g}\right\},\\
 \label{eq:matrixT_Main_results}
 \mathcal{\check{T}}^{S}&= -\frac{i}{2}\alpha n_r \epsilon_{kjr}\left[\sigma_{j},\ \check{g} \tilde{\nabla}_k\check{g} \right]  \nonumber\\
 &+\frac{i}{8}\gamma n_{r}\epsilon_{ijl}\epsilon_{lkr}\left[\left\{ \tilde{\nabla}_k \check{g},\ \check{g}\sigma_{i}\check{g}\right\},\ \sigma_{j}\right]\;,
\end{align}
where we keep terms up to second order in Pauli matrices, reflecting that SOC is usually weak.
The detailed derivation of the above boundary condition is presented in 
 Sec.~\ref{sec:sigma_model}. 

Equations (\ref{eq:Boundary_ISOC_main}-\ref{eq:matrixT_Main_results}) constitute one of the main results of this work.  
They describe, for example,  the electrical and spin surface currents that may appear due to the ISOC: 
\begin{subequations}
\begin{align}
    \vec{J}^{S} &= -\frac{e\pi\nu_{0}}{4}\text{tr}\{\check{\rho}_{1}\check{\tau}_{3}\vec{\mathcal{\check{J}}}^{S}\}\;,\\
    \vec{j}^{S}_{a} &= -\frac{\pi\nu_{0}}{4}\text{tr}\{\check{\rho}_{1}\sigma_{a}\vec{\mathcal{\check{J}}}^{S}\}\;,
\end{align}
\end{subequations}
where $\check{\rho}_{i}$ and $\check{\tau}_{j}$ stand for the Pauli matrices in Keldysh and Nambu spaces, respectively.
Indeed, Eq.~(\ref{eq:Boundary_ISOC_main}) describes three types of effects: spin-precession characterized by the coefficient $\alpha$, the spin-galvanic term characterized by $\gamma$ and the spin-relaxation terms governed by the tensor $\Gamma _{ab}$, which for an isotropic interface has only two different components: $\Gamma^{\parallel}$ and $\Gamma^{\perp}$, see Eq.~(\ref{eq:Relaxationtensor}).  

For illustrative purposes, let us analyze the normal state.  In this case, see Appendix~\ref{sec:NormalState} for details, from  Eq.~(\ref{eq:Boundary_ISOC_main}) one obtains the following boundary conditions for the charge and spin components:
\begin{align}
\label{eq:Normal_state_main_results2}
    D \vec{n}\cdot\vec{\nabla} \delta n &= -\gamma \vec{n}\times\vec{\nabla}\cdot \delta\vec{S}\;,\\
    D\vec{n}\cdot\vec{\nabla} \delta \vec{S}^{\parallel} &= -\gamma \vec{n}\times\vec{\nabla}\delta n + 8\Gamma^{\parallel} \delta \vec{S}^{\parallel}\nonumber\\& - 2\alpha (\vec{n}\times\vec{\nabla})\times\vec{n}\delta S^{\perp}\;,\\
    D\vec{n}\cdot\vec{\nabla}\delta S^{\perp}&=8\Gamma^{\perp}\delta S^{\perp}-2\alpha \vec{n}\cdot\left((\vec{n}\times\vec{\nabla})\times\delta\vec{S}\right)\label{eq:Normal_state_main_results2perpendicularspin}\;.
\end{align}
Here $\delta n$ and $\delta \vec{S}$ are the excess charge and the excess spin in the material and for simplicity of notation we defined $S^{\perp} = \vec{n}\cdot\vec{S}$ and $\vec{S}^{\parallel} = \vec{S}-S^{\perp}\vec{n}$. These expressions coincide with  the boundary conditions derived in \cite{borge2019boundary}.
By comparison, $\Gamma^{\parallel,\perp}$  can be identified with the spin loss coefficients $L^{\parallel,\perp}$ and spin loss conductances $G^{\parallel,\perp}$ via $\Gamma^{\parallel,\perp} = L^{\parallel,\perp}/8 = G^{\parallel,\perp}/(8e^{2}\nu_{0})$, while $\gamma$ can be related to the spin-charge conductivity $\sigma^{sc}$ of the interface \cite{sanz2020quantification} via $\gamma = -D{\sigma^{sc}}/{\sigma_{D}} = -{\sigma^{sc}}/{2e^{2}\nu_{0}}$, where $\sigma_{D}$ is the normal state conductivity of the superconductor, and $\nu_0$ the density of states at the Fermi level. Meanwhile, $\alpha$ is related to the spin-swapping coefficient $\kappa$ via $\alpha = -D\kappa/2$. Thus, all parameters  can be uniquely determined by considering the normal state.

Eq.~(\ref{eq:Boundary_ISOC_main}) has a broader range of applicability, capturing all symmetry-allowed effects at the interface, not only in the normal state but also in the superconducting state.  
To illustrate this and highlight how the new boundary conditions smoothly connect the normal and superconducting states, we consider a simple limiting case for temperatures just below the superconducting critical temperature. We focus on an equilibrium situation and therefore we only focus on the retarded component of $\check g$, denoted here by the same symbol.  In the linearized  regime, it can be written as $\check  g= \delta(t-t')\check{\tau}_{3}+\check f$ where the pair amplitude matrix $\check f$ is a small correction. The general  matrix structure in Nambu-spin space  reads:
\begin{align}
    \check{g} = \delta(t-t')\check{\tau}_{3}+\begin{pmatrix}
    0&f_{s}+\vec{f}_{t}\cdot \vec{\sigma}\\-\Tilde{f}_{s}-\Tilde{\vec{f}}_{t}\cdot \vec{\sigma}&0
\end{pmatrix}\;.
\end{align}
Linearization of  Eq.~(\ref{eq:Boundary_ISOC_main}) leads to the following boundary conditions:
\begin{align}
\label{eq:main_linearized1}
    D \vec{n}\cdot\vec{\nabla} f_{s} &= - \gamma \vec{n}\times\vec{\nabla}\cdot \vec{f}_{t}\;,\\
    D\vec{n}\cdot\vec{\nabla}\vec{f}_{t}^{\parallel} &=  - \gamma \vec{n}\times\vec{\nabla}f_{s} + 8\Gamma^{\parallel}  \vec{f}_{t}^{\parallel}\nonumber\\& - 2\alpha (\vec{n}\times\vec{\nabla})\times\vec{n} f_{t}^{\perp}\;,\\
    D\vec{n}\cdot\vec{\nabla} f_{t}^{\perp}&= 8\Gamma^{\perp}f_{t}^{\perp}-2\alpha \vec{n}\cdot\left((\vec{n}\times\vec{\nabla})\times\vec{f}_{t}\right)\;,
    \label{eq:main_linearized3}
\end{align}
and identical expressions  for $\tilde{f}_{s},\tilde{\vec{f}}_{t}$.

Comparison of Eqs.~(\ref{eq:main_linearized1})-(\ref{eq:main_linearized3}) with Eqs.~(\ref{eq:Normal_state_main_results2}-\ref{eq:Normal_state_main_results2perpendicularspin}) reveals the connection between charge-spin densities and singlet-triplet condensates. Specifically, these equations become identical under the replacement $\delta n\xrightarrow{}f_{s}$ and $\delta\vec{S}\xrightarrow{}\vec{f}_{t}$.
This connection, previously noted at the level of the Usadel equation in Ref.~\cite{bergeret2014spin}, arises from the fact that spin-orbit coupling does not break time-reversal symmetry.

From this analogy, we conclude that in the superconducting state, $\Gamma^{\parallel,\perp}$  describe triplet relaxation, $\gamma$ leads to interfacial singlet-triplet conversion, while $\alpha$ mixes different types of triplets. 
Consequences of these terms, such as a modification of the critical temperature, or  increase of the superfluid stiffness, are explored in Sec.~\ref{sec:GL}. Well below the critical temperature they lead to the appearance of the superconducting diode effect, which, as discussed in \ref{sec:Diode}, can be large if singlet-triplet conversion dominates over triplet relaxation. In the next  section, we derive the boundary conditions presented in Eqs.~(\ref{eq:matrixJ_Main_Results}) and (\ref{eq:matrixT_Main_results})
from  a quasiclassical boundary action.

\section{Derivation of the effective action}
\label{sec:sigma_model}

\begin{figure}[htbp]
    \centering
    \includegraphics[width=0.9\linewidth]{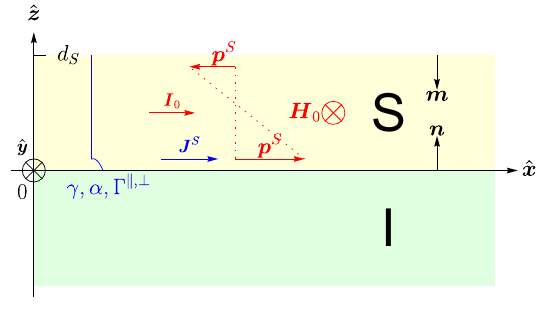}
    \caption{The system under consideration. We show schematically the distribution of the superconducting momentum, $\vec{p}^S = \nabla \varphi - \frac{2e}{c}\vec{A}$, across the film caused by the external magnetic field $\vec{H}_0$ and the charge current $\vec{I}_0$. $\vec{n}$ and $\vec{m}$ are the unit vectors perpendicular to the interfaces at $z = 0$, $z = d_S$, respectively.  The spin-galvanic coefficient $\gamma$, the spin precession coefficient $\alpha$, and the spin-relaxation terms $\Gamma^{\parallel}$ and $\Gamma^{\perp}$ exist only at the interface with a heavy atom insulator at $z=0$.}
    \label{fig:setup}
\end{figure}

To derive the equations presented in the preceding section, we
consider the generic setup shown in Fig.~\ref{fig:setup}. 
A superconducting material extends infinitely in the $\vec{x}-\vec{y}$ plane and has thickness $d_{S}$. It lies on top of an insulating substrate that occupies the region $z<0$.  If the insulator contains heavy atoms, then the interface between the superconductor and the insulator allows for a sizable interfacial spin-charge coupling at $z = 0$. The top surface of the superconductor,  at $z = d_{S}$,  interfaces with vacuum \footnote{Alternatively, another insulator may be present for $z>d_{S}$ region. However, that insulator must differ from the bottom one; otherwise, the inversion symmetry is restored and the system is not gyrotropic anymore. }. Due to the asymmetry between the top and bottom surfaces, the system is polar. For this reason, the symmetry breaking requirements for spin-galvanic effects are met \cite{kokkeler2024nonreciprocal,kokkeler2024universal}: anomalous currents may be driven by an external magnetic field $\vec{H}_{0}$, and vice versa, spin accumulation may be generated by the application of an external current $\vec{I}_{0}$. 

To investigate whether such effects appear and to assess how strong they are, we derive the quantum transport equations in the diffusive limit. These equations can be obtained with the help of the so-called nonlinear sigma model (NLSM), which describes the soft-diffusion modes via a generating functional 
\begin{equation}
    Z\left[\left\{V_{i}\right\}\right] = \int D\left[\check{Q}\right] e^{iS\left[\check{Q},\ \left\{V_{i}\right\} \right]}\;,\label{eq:Generatingfunctional}
\end{equation}
that depends on the collection of external fields $\{V_{i}\}$ and on the matrix field $\check{Q}$, whose expectation value is the quasiclassical Green's function (GF) $\check{g}$. It can be shown that if the Fermi level and the scattering frequency are the largest energy scales of the system, the main contribution to the generating functional stems from the so-called soft-mode manifold of field configurations that satisfy the nonlinear constraint $\check{Q}^{2} = \check{1}$. For more details, we refer to the extensive literature on this topic; see, e.g., Refs.~\cite{feigelman2000keldysh, kamenev2023field} and the references therein.

Any physical observable can be found by calculating the derivative of the generating functional Eq.~(\ref{eq:Generatingfunctional}) with respect to an appropriate field $V_{i}$. This means that an observable can be expressed as a weighted average over the different $\check{Q}$-field configurations. Since the action $S\left[\check{Q},\ \left\{V_i\right\}\right]$ appears as the weighting factor for these configurations, a good approximation is given by taking into account only the contribution of the saddle point configuration.
Specifically, applying this consideration to the expectation value of $\check{Q}$, we identify the quasiclassical Green's function with the saddle point configuration of $\check{Q}$. This means that the saddle point condition of the action of the NLSM gives the governing equation for $\check{g}$, that is, it forms the quantum transport equation, which in the dirty limit is called the Usadel equation.




In line with previous works on NLSMs for superconductors with spin-orbit coupling \cite{virtanen2021magnetoelectric,virtanen2022nonlinear,kokkeler2024universal}, we use the Keldysh version of the NLSM because of its applicability to a wide range of systems \cite{feigelman2000keldysh,kamenev2023field}. In the Keldysh formalism, in order to make the calculation of observables from the generating function tractable, one chooses a contour in time space that contains two branches, one that runs from $-\infty$ to $\infty$ and one in the opposite direction \cite{Keldysh1965diagram,van2006introduction,feigelman2000keldysh,kamenev2023field}. These branches can be described with the help of an additional matrix space spanned along Pauli matrices $\check{\rho}_{i}, i = 1,2,3$ and $\check{\rho}_0 = \check{1}$ \cite{kamenev2023field}. There are several possible choices for this construction; here we use the one in which causal functions take an upper triangular form \cite{kamenev2023field}.  
Here and hereafter, we use the check notation $\check{\cdot}$ to specify matrices in Nambu and Keldysh spaces. $\check{\tau}_{1,2,3}$ and $\sigma_{x,y,z}$ are the Pauli matrices in Nambu and spin spaces, respectively. $\vec{\sigma} = (\sigma_x,\ \sigma_y,\ \sigma_z)$. 

The bulk action of the superconductor can be written within the NLSM approach as a functional over the matrix field $\check{Q}$ that satisfies the constraint $\check{Q}^2 = \check{1}$, the matrix order parameter $\check{\Delta}$, and the electromagnetic matrix vector potential $\check{\vec{A}}$. Because the pair potential and the vector potential may take different values on the two time branches, they contain two components in Keldysh space, the classical  
components $\vec{A}\check{\rho}_{0},\ \Delta e^{i\varphi \check{\tau}_{3}}\check{\tau}_{2} \check{\rho}_{0}$ which correspond to the fields, i.e. the vector potential $\vec{A}$ and the magnitude $\Delta$ and the phase $\varphi$ of the pair potential, taking the same value on both time branches, and the quantum components $\vec{A}^{q}\check{\rho}_{1},\ \Delta^{q} e^{i\varphi^{q}\check{\tau}_{3}}{\check{\tau}_{2}}\check{\rho}_{1}$ which correspond to these fields that take different values on both time branches
\cite{,kamenev2023field,kokkeler2024nonreciprocal,kokkeler2024universal}. 
The classical and quantum components together form the pair potential $\check{\Delta} = \Delta e^{i\varphi \check{\tau}_{3}}\check{\tau}_{2}\check{\rho}_{0}+\Delta^{q}e^{i\varphi^{q} \check{\tau}_{3}}\check{\tau}_{2}\check{\rho}_{1}$, and the vector potential $\check{\vec{A}} = \vec{A}\check{\rho}_{0}+\vec{A}^{q}\check{\rho}_{1}$ in Keldysh space. 

As discussed in \cite{kokkeler2024nonreciprocal}, all terms in the action can be constructed by taking into account two main symmetries of the internal structure of the NLSM and the crystal symmetries of the material. First of all, by construction of Nambu space, all the terms in the action need to satisfy the so-called charge conjugation symmetry, which fixes the allowed terms and the structure of the corresponding tensors. Physically, charge conjugation symmetry means that the system's behavior remains the same whether it is described using electron or hole representations. Secondly, the terms have to satisfy the chronological symmetry, which reflects  the causality,  and determines whether the coefficients are real or imaginary, and thereby guarantees that all the observables are real.

For a cubic bulk normal metal without spin-orbit coupling, the terms in the action obeying these symmetries up to the first order in spin Pauli matrices are well known and read \cite{feigelman2000keldysh}, 
\begin{equation}\label{eq:Bulk_Action}
    \begin{aligned}
     &iS_{\text{bulk}}\left[\check{Q},\ \check{\Delta},\ \check{\vec{A}}\right] = \int dV  \operatorname{Tr}\Big\{\frac{\pi \nu_0}{2}\Big( - \frac{ i}{2\pi \lambda} \check{\Delta}^\dagger \check{\rho}_1\check{\Delta}- \\ 
     &\quad -\frac{D}{4}(\tilde{\nabla}_i \check{Q})^{2} +\hat{\omega}_{t,t'}\check{\tau}_{3}\check{Q}  
     +\check{\Delta}\check{Q}+i(\vec{h}\cdot\vec{\sigma})\check{\tau}_{3}\check{Q} \Big) \\
     &\quad -i \left[  \frac{\check{\rho}_{1}(\nabla\times\check{\vec{A}})^{2}}{8\pi} - \frac{\check{\rho}_{1}\nabla\times \check{\vec{A}} \cdot \vec{H}_0}{4 \pi} \right]   \Big\}\;.
\end{aligned}
\end{equation}
The trace $\operatorname{Tr}$ is taken over Keldysh-Nambu-spin space and time, while the spatial integration is performed over the entire space, where it is taken into account that the $\check{Q}$-matrices are finite only within the superconductor. $\nu_0$ is the normal density of states of the superconductor, $\lambda > 0$ represents the BCS coupling constant, $D$ is the diffusion coefficient, and $\hat{\omega}_{t,t'} = \delta(t-t')\partial_{t}$ represents the time derivative.   The vector potential is incorporated into the gauge invariant derivative, $\tilde{\vec{\nabla}} = \vec{\nabla}-\left[\frac{ie}{c}\vec{\check{A}}\check{\tau}_{3}, \cdot\right]$, with $c$ and $e$ being the speed of light and the charge of free electrons, respectively. $\vec{h}$ is the Zeeman field in the superconductor. In dirty materials, as discussed in Appendix~\ref{sec:Maxwell}, it is related to the external magnetic field via 
\begin{align}\label{eq:ExchangefromExternal}
\vec{h} = g_L \mu_B \vec{H}_0\;,
\end{align}
where $g_L$ and $\mu_B$ represent the Landé factor and the Bohr magneton, respectively. 
The magnitude and the phase of the pair potential are denoted as $\Delta$ and $\varphi$ 
respectively. They must be determined using selfconsistency by requiring the action to also be a saddle point with respect to $\check{\Delta}$. The first Pauli matrix $\check{\rho}_{1}$ appears in terms that do not involve $\check{Q}$ since, due to the opposite orientations of the two branches, the contributions of these branches to the action must be subtracted. 


The action in Eq.~(\ref{eq:Bulk_Action}) is the standard bulk action for a superconductor in a magnetic field and does not describe any spin-orbit coupling effects, which in principle are always present. However, if we assume that the crystal of the metal is inversion symmetric, as is the case for many metals, the only terms that may arise are the spin-Hall and spin-swapping terms, which are usually small in superconductors with small atomic numbers such as aluminum or niobium. Therefore, for illustrative purposes, we will ignore such terms and instead focus on interfacial spin-orbit coupling.



This interfacial spin-orbit coupling may in principle appear at both boundaries. However, for boundaries with vacuum, interfacial spin-orbit coupling is usually small in materials with small spin-Hall and spin-swapping coefficients and therefore we neglect it here.   
On the other hand, in  insulators with  heavy elements such as  bismuth or tungsten  \cite{sanz2020quantification,Groen23}, interfacial spin-orbit coupling at this boundary can be large compared to bulk spin-orbit coupling.  Because interfacial spin-charge coupling is relativistic, we may use an effective action describing such boundaries by constructing all symmetry allowed scalars composed of the matrix field $\check{Q}$, spatial derivatives and spin-Pauli matrices, up to second order in spin, similar to the recent approach in Ref.~\cite{kokkeler2024universal}. Moreover, since we consider a boundary action for a second-order differential equation, we keep terms up to first order in spatial derivatives. Keeping in mind the normalization condition $\check{Q}^{2} = \check{1}$, we construct all possible terms allowed by the charge conjugation symmetry introduced in Ref.~\cite{kokkeler2024universal}:
\begin{align}\label{eq:SOC_action}
    iS_{SOC}\left[\check{Q},\ \check{\vec{A}}\right] = \int_{\mathcal{S - I}} dS\ \frac{\pi \nu_0}{2} \operatorname{Tr}\Big\{\frac{i}{2 }\alpha_{kj}\sigma_{j}\check{Q} \tilde{\nabla}_{k}\check{Q} \nonumber\\
    -\frac{i}{8}\gamma_{lk} \epsilon_{ijl}\sigma_{i}\check{Q}\sigma_{j}\check{Q}\tilde{\nabla}_{k}\check{Q} + \Gamma_{ab}\sigma_{a}\check{Q}\sigma_{b}\check{Q}\Big\}\;,
\end{align}
where $\mathcal{S - I}$ is the boundary between the superconductor and the insulator. The first of these terms is the spin-precession term, and can be included as an SU(2) covariant derivative \cite{virtanen2022nonlinear}, while the second term is responsible for the spin-galvanic effect \cite{ganichev2002spin,virtanen2022nonlinear}. Both terms require the existence of a second rank pseudotensor, that is, a gyrotropic tensor $\gamma_{lk}$. For boundaries, the condition of gyrotropy is  satisfied, because the normal to the interface is a polar vector of the system \cite{kokkeler2024nonreciprocal}. Consequently, for an isotropic boundary the spin-precession tensor and spin-galvanic tensor need to take the form \begin{align}
    \alpha n_{r}\epsilon_{kjr}\;,\label{eq:alphastructure}\\
    \gamma n_{r}\epsilon_{lkr}\;,\label{eq:gammastructure}
\end{align} 
that is, they have the same structure as for SU(2) covariant Rashba systems \cite{virtanen2022nonlinear}. As shown in Appendix \ref{sec:NormalState} with the help of the normal state equations, $e^{2}\nu_{0}\gamma$ can be identified as the Edelstein conductivity, while $2D\alpha$ corresponds to the spin-swapping coefficient in \cite{borge2019boundary}. The third term describes spin relaxation. The spin-relaxation tensor $\Gamma_{ab}$ may have contributions from many different processes, such as Dyakonov-Perel \cite{Dyakonov1971} or Elliott-Yafet \cite{elliot1954theory,Yafet1952} mechanisms or electron-electron correlations \cite{glazov2004effect}. The spin-relaxation tensor is a second-rank symmetric tensor, and may exist for any crystal symmetry. For an isotropic boundary with normal vector $\vec{n}$,  it takes the form, 
\begin{align}\label{eq:Relaxationtensor}
\Gamma_{ab} = \frac{1}{2}\Gamma^{\perp}(\delta_{ab}-2n_{a}n_{b})+\Gamma^{\parallel} n_{a}n_{b}\;,
\end{align} where $\Gamma^{\parallel}$ denotes the spin relaxation for in-plane spins and $\Gamma^{\perp}$ the relaxation for out-of-plane spins. All these tensors $\alpha_{kj}$, $\gamma_{lk}$ and $\Gamma_{ab}$ are real by the chronology symmetry \cite{kokkeler2024universal}.


Introducing the forms of the tensors, Eqs.~(\ref{eq:alphastructure}--\ref{eq:Relaxationtensor}) in Eq.~(\ref{eq:SOC_action}) we obtain the following boundary action for our theory:
\begin{align}\label{eq:Sboundary}
    &iS_{B}\left[\check{Q},\ \vec{A}\right] = \int_{\mathcal{S-I}} dS\ \frac{\pi\nu_{0}}{2}\operatorname{Tr}\Big\{-\frac{i\alpha}{2}\vec{n}\times \vec{\sigma}\cdot \check{Q}\tilde{\nabla}\check{Q}  \nonumber\\
    &\ + \frac{1}{2}\Gamma^{\perp}\vec{\sigma}_\parallel\cdot \check{Q}  \vec{\sigma}_\parallel 
    \check{Q} + \left(\Gamma^{\parallel}-\frac{1}{2}\Gamma^{\perp}\right)(\vec{n}\cdot \vec{\sigma}) \check{Q}(\vec{n}\cdot\vec{\sigma})\check{Q}\Big\}\nonumber\\
    &\ -\frac{i\gamma}{8}\check{Q}\vec{n}\cdot \vec{\sigma} \check{Q} \left\{\vec{\sigma},\ \tilde{\nabla} \check{Q} \right\} \;. 
\end{align}
Here, $\vec{\sigma}_\parallel = \vec{\sigma} - (\vec{n}\cdot\vec{\sigma}) \vec{n}$.

The action. Eq.~(\ref{eq:Sboundary}) is a central result of our work. It is the boundary action for any non-magnetic isotropic boundary between a superconductor and an insulator up to second order in the relativistic spin-orbit coupling, valid for any temperature. According to it, such boundaries are described by four independent quantities, $\alpha,\gamma,\Gamma^{\parallel}$ and $\Gamma^{\perp}$. Generally, within our formalism there is no specific relation between the four independent coefficients, which may depend in an intricate way on the band structure and the type of spin-orbit coupling.  Parameters like  $\gamma$ and $\Gamma^{\parallel,\perp}$  can be extracted from experiments in the normal state \cite{sanz2020quantification,pham2021large,Groen23}.  



The total action functional of the superconductor is given by the sum of the bulk, Eq.~(\ref{eq:Bulk_Action}),  and boundary, Eq.  (\ref{eq:Sboundary}), contributions 
\begin{equation}
    \label{Eq.)TOTAL_action}
S\left[\check{Q},\ \check{\Delta},\ \check{\vec{A}}\right] = S_{\text{bulk}}\left[\check{Q},\ \check{\Delta},\ \check{\vec{A}}\right] + S_{B}\left[\check{Q},\ \check{\vec{A}}\right]\,.
\end{equation}
To derive the transport equations  we use the saddle point approximation. 
At the saddle point, 
the $\check{Q}$-matrix can be identified with the quasiclassical Green's function $\check{g}$, which has the usual causality structure, 
\begin{align}
    \check{g} = \begin{pmatrix}
        \check{g}^{R}&\check{g}^{K}\\0&\check{g}^{A}
    \end{pmatrix}\;.
\end{align}
The saddle point equations can be determined by requiring that the functional derivatives with respect to  these quantities have to vanish. First of all, we require the invariance under variations of the $\check{Q}$-matrix, $\frac{\delta S}{\delta \check{Q}} = 0$, which leads to the Usadel equation: 
\begin{equation} 
 \label{eq:Usadel}
   D\tilde{\nabla}_{k}(\check{g}\tilde{\nabla}_{k} \check{g}) = \left[\hat{\omega}_{t,t'} \check{\tau}_{3}+  \check{\Delta}+i(\vec{h}\cdot\vec{\sigma})\check{\tau}_{3},\ \check{g}\right]\; ,
\end{equation}
and the boundary condition at the interface with SOC:
   \begin{equation}
    \left.-D \check{g}\tilde{\nabla}_{z} \check{g} \right\vert_{z = 0} =  
 -\vec{\tilde{\nabla}}\cdot \vec{\mathcal{\check{J}}}^{S}+\mathcal{\check{T}}^{S}-2\Gamma_{ab}\left[\sigma_{a}\check{g}\sigma_{b},\ \check{g}\right]\;.   \label{eq:Boundary_ISOC_main_1}
\end{equation}
 The terms on the right-hand side originate from the boundary action, Eq.~(\ref{eq:Sboundary}), where $\vec{\mathcal{\check{J}}}^{S}$ represents the matrix surface current and $\mathcal{\check{T}}^{S}$ is the matrix surface torque, defined as:
\begin{subequations}
    \begin{align}
     \label{eq:matrixJ}
 \mathcal{\check{J}}^{S}_{k} &= \frac{i}{2} \alpha n_r \epsilon_{kjr} \left[\check{g}\sigma_{j},\ \check{g}\right] + \nonumber\\&+\frac{i}{16}\gamma n_{r}\epsilon_{ijl}\epsilon_{lkr}\left\{\left[\sigma_{i},\ \check{g}\right],\ \sigma_{j}+\check{g}\sigma_{j}\check{g}\right\}\;,\\
 \label{eq:matrixT}
 \mathcal{\check{T}}^{S}&= -\frac{i}{2}\alpha n_r \epsilon_{kjr}\left[\sigma_{j},\ \check{g} \tilde{\nabla}_k\check{g} \right]  \nonumber\\
 &+\frac{i}{8}\gamma n_{r}\epsilon_{ijl}\epsilon_{lkr}\left[\left\{ \tilde{\nabla}_k \check{g},\ \check{g}\sigma_{i}\check{g}\right\},\ \sigma_{j}\right]\;.
\end{align}
\end{subequations}
In Appendix~\ref{sec:NormalState} we show that in the normal state the boundary condition, Eq.~(\ref{eq:Boundary_ISOC_main_1}), coincides with the one derived in  Ref.~\cite{borge2019boundary}. In particular, the term proportional to $\gamma$ describes the spin-galvanic effect, the $\alpha$ term corresponds to the spin precession, and terms $\Gamma^{\parallel,\perp}$ describe the spin-relaxation. 

We note that the boundary conditions for the Usadel equation, Eq.(\ref{eq:Boundary_ISOC_main_1}), differ from those obtained in Refs.~\cite{Cottet09boundary,Eschrig_2015}. In fact, Eq. 
(\ref{eq:Boundary_ISOC_main_1}) is formulated specifically for interfaces with non-magnetic heavy-atom insulators, whereas Refs.~\cite{Cottet09boundary,Eschrig_2015} focus solely on magnetic interfaces. In contrast to previous works, our boundary conditions incorporate ISOC which leads to a spin-dependent mixing of in-plane and normal derivatives, and generates the spin-galvanic effect.

Next, we consider the variation of the action Eq.~(\ref{Eq.)TOTAL_action}) with respect to $\check{\Delta}$. We know that if there are no quantum components, that is, the fields take the same value on both branches, the contributions of the two branches cancel out each other exactly, and hence the action vanishes \cite{kamenev2023field}. Thus, the saddle point conditions are satisfied by requiring that at the saddle point the quantum components vanish, i.e. ${\Delta}^{q} = 0$, $\varphi^{q} = 0$, which automatically guarantees that $\frac{\delta S}{\delta {\Delta}} = 0$, while additionally imposing $\frac{\delta S}{\delta {\Delta}^{q}} = \frac{\delta S}{\delta {\varphi}^{q}} = 0$. The latter condition results in the well-known self-consistency relation for the magnitude of the pair potential $\Delta$ \cite{kamenev2023field} and its phase $\varphi$:
\begin{align}\label{eq:Selfconsistency}
    \Delta e^{i\varphi} =\frac{\pi\lambda}{8}\operatorname{tr}\Big(\check{\rho}_{1}(\check{\tau}_{1}-i\check{\tau}_{2})\check{g}(t,t)\Big)\;,
\end{align}
where tr denotes tracing over matrix indices in Nambu-Keldysh-spin space, but without time integration.

We finally take the functional derivative of Eq.~(\ref{Eq.)TOTAL_action}) with respect to the vector potential.  The saddle point condition is satisfied by requiring that at the saddle point $\vec{A}^{q} = 0$  and $\frac{\delta S}{\delta \vec{A}^{q}} = 0$, which leads  to the Maxwell equation for the vector potential $\vec{A}$:
\begin{equation}
\label{eq:Maxwell}
\nabla \times \left[\nabla\times \vec{A}\right]= \frac{4\pi}{c}\vec{j}_S\; , 
\end{equation}
where  $j_{S,k} = \frac{ e\pi\nu_0 D}{4}  \operatorname{tr}\left\{  \check{\rho}_{1}\check{\tau}_{3} \check{g}\tilde{\nabla}_{k}\check{g} \right\}$ is the supercurrent density distributed in the film. 
The  boundary conditions for the Maxwell equation at the film edges are: 
\begin{subequations} \label{eq:Maxwell_BCs}
    \begin{align}
    & \vec{m}\times \Big(\nabla\times \vec{A}(d_S-0) - \vec{H}_{0}\Big) = -\frac{2\pi}{c}\vec{I}_{0}\;,\\
    & \vec{n} \times \Big(\nabla\times\vec{A}(+0) -\vec{H}_{0}\Big) = \frac{4\pi}{c} \vec{J}^S-\frac{2\pi}{c}\vec{I}_{0}\; .
\end{align}
\end{subequations}
Here, $\vec{n} = (0,\ 0,\ 1)$ and $\vec{m} = (0,\ 0,\ -1)$ are the unit vectors perpendicular to the interfaces at $z = 0$ and $z = d_S$ respectively.  $J^{S}_{k} = -\frac{e\pi\nu_{0}}{4}\operatorname{tr} \left\{ \check{\rho}_1 \check{\tau}_{3}\mathcal{\check{J}}^{S}_{k} \right\}$ is the surface anomalous current due to the ISOC, where $\mathcal{\check{J}}^{S}_{k}$ is defined in Eq.~(\ref{eq:matrixJ}).  
The term $\vec{I}_{0}$ is usually imposed to fix the total current in the wire by fixing the magnetic field produced by the wire. It is given by  $\vec{J}^{S} + \int dz \vec{j}_{S} = \vec{I}_{0}$. The appearance of this term in Eqs.~(\ref{eq:Maxwell_BCs})  can be derived within the action formalism by adding a term $-\frac{1}{2 c}\check{\rho}_{1}\vec{\check{I}}_{0}\cdot\check{\vec{A}}$ to the action of  both boundaries, Eq.  (\ref{eq:Sboundary}), and zero action at the vacuum boundary.


Eqs.~(\ref{eq:Usadel}--\ref{eq:Maxwell_BCs}) is a complete set of equations determining  the spectral and transport properties of superconducting systems with interfacial spin-orbit coupling in the presence of external fields and currents 
In particular, the boundary condition, Eq.~(\ref{eq:Boundary_ISOC_main_1}) is  one of the main results of this work, generalizing the boundary conditions for ISOC in the normal state obtained in Ref.~\cite{borge2019boundary} to superconductors. In the next section, we introduce some approximations that will be used throughout the rest of the article to compute the critical temperature, anomalous current, and diode effect of thin superconducting films with ISOC. 

\subsection{Thin superconducting films}
As stated above, the boundary value problem posed in Eqs.~(\ref{eq:Usadel}--\ref{eq:Maxwell_BCs}) describes the system in both normal and superconducting states, and for generic orientations and spatial dependence of the magnetic field and the applied current in the plane. In the following sections, we focus on  equilibrium  properties of the superconducting state. For this reason, it is convenient to use the Matsubara representation instead of the full Keldysh one.
This means that in all previous equations we have to make the substitution $\hat{\omega}_{t,t'}\xrightarrow{}\omega_{n} = (2n+1)\pi T$, where $n$ is an integer. 
Moreover, it is convenient to solve Eqs.~(\ref{eq:Usadel}--\ref{eq:Selfconsistency}) in a gauge in which the superconducting phase $\varphi$ in $\check{\Delta} = \Delta e^{i\check{\tau}_3 \varphi} \check{\tau}_2$ is eliminated. This is achieved via a gauge transformation which leads to the following transformations:
\begin{equation} \label{eq:Phase_Rotation_Delta}
\begin{aligned}
    &\check{g} \to  e^{i\varphi \check{\tau}_3/2} \check{g}(z)  e^{-i\varphi \check{\tau}_3/2}\;,\\
    &\check{\Delta} \to e^{-i\varphi \check{\tau}_3/2} \check{\Delta}  e^{i\varphi \check{\tau}_3/2} =  \Delta  \check{\tau}_2,\\
    &\Tilde{\nabla}_{\parallel} \xrightarrow{}\frac{i\vec{p}^{S}}{2}\left[\check{\tau}_{3}\;,\ \check{\cdot}\right],\qquad
    \tilde{\nabla}_{z}\xrightarrow{}\nabla_{z}\; , 
\end{aligned}
\end{equation} 
where $\vec{p}^S = \nabla \varphi - \frac{2e}{c}\vec{A}.$
After this  gauge transformation, instead of two gauge-dependent quantities $\varphi$ and $\vec{A}$, we end up with only the superconducting momentum $\vec{p}_S$ that is gauge invariant.  

To maximize the effects of ISOC, we assume that the externally applied magnetic field and the current are homogeneous and that the magnetic field is in-plane. We define this direction as the $\hat{\vec{y}}$-direction, i.e. $\vec{H}_{0} = H_{0}\vec{\hat{y}}$, so that by symmetry of the spin-galvanic tensor in Eq.~(\ref{eq:gammastructure}) the anomalous current flows in the $\hat{\vec{x}}$-direction. Therefore, the externally imposed current is always chosen as $\vec{I}_{0} = I_{0}\vec{\hat{x}}$. In this case, the symmetries of the Usadel equation imply that $\vec{p}^{S} = p^{S}\vec{\hat{x}}$ and $\check{g} = \check{g}_{s}+\check{g}_{t}\sigma_{y}$. Following Eqs.~(\ref{eq:matrixJ}) and (\ref{eq:matrixT}) this has the consequence that all terms that depend on $\alpha$ drop out of the equation. Indeed, in the normal state this leads to precession of spins around the magnetic field axis, while in the superconducting state it leads to a precession of the zero-spin polarization axis. In the setup here the zero-spin polarization axis is along the magnetic field and therefore does not precess. 


Since interfacial effects are most prominent in thin films, we focus on a superconducting film with a thickness  much smaller than the coherence length. Namely,  we assume that 
\begin{equation}
    d_S \ll \xi = \sqrt{\frac{D}{2\pi T_{c0}}}\;.    \label{eq:Approximation_1}
\end{equation}
Here $T_{c0}$ is the BCS critical temperature corresponding to the coupling constant $\lambda$ written in Eq.~(\ref{eq:Bulk_Action}). Moreover, since we study  the diffusive regime, the coherence length $\xi$ is much shorter than the London penetration depth $\lambda_{L}$,
\begin{equation}
    \xi \ll \lambda_L\;. \label{eq:Approximation_2}
\end{equation}
As derived in Appendix \ref{sec:Maxwell}, when Eq.~(\ref{eq:Approximation_2}) is satisfied, the Maxwell equations in Eq.~(\ref{eq:Maxwell}--\ref{eq:Maxwell_BCs})  impose that the magnitude $p^{S}$ of the superconducting momentum reads 
\begin{align} \label{eq:pS_profile}
    p^{S}(z) = p^{S0}-\frac{2eH_{0}}{c}(z-d_{S}/2)\;.
\end{align} 
Here, $p^{S0}$ is the uniform part of the superconducting momentum. Within our calculations, we fix $p^{S0}$, solve the Usadel equation, find $\Delta$, and then we compute the total current corresponding to the given $p^{S0}$ as
\begin{equation}\label{eq:I0Eqmaintext}
    I_0 = W p^{S0} d_S + J^S\;, 
\end{equation}
so $p^{S0}$ depends on the anomalous current $J^S$, the applied current $I_0$, and the superconducting stiffness $W$. The latter is  given by
\begin{equation}\label{eq:superconductingstiffnessMainText}
    W = - \frac{e \pi \nu_0 D T}{4} \sum_{\omega_n} \operatorname{tr}\left\{\check{\tau}_3 \check{g} \left[\check{\tau}_3,\ \check{g}\right]\right\}\;.     
\end{equation} 

Using the approximation, Eq.~(\ref{eq:Approximation_1}), and the superconducting momentum $p^{S}(z)$ from Eq.~(\ref{eq:pS_profile}), we average the Usadel equation Eq.~(\ref{eq:Usadel}) over the thickness of the superconductor and obtain the following algebraic relation for $\check{g}$: 

\begin{align}
    &-\frac{d_S D (p^{S0})^{2}}{4}\left[ \check{\tau}_{3},\ \check{g}\left[\check{\tau}_{3},\ \check{g}\right]\right]  \nonumber \\ 
    &- \frac{i}{2}\left(p^{S0}+\frac{eH_{0}d_{S}}{c}\right)\left[\check{\tau}_{3},\ \mathcal{\check{J}}^{S}_x\right]  +\mathcal{\check{T}}^{S} = \nonumber \\
    & = d_S\left[\omega_{n}\check{\tau}_{3}+\Delta\check{\tau}_{2}+i(\vec{h}\cdot\vec{\sigma})\check{\tau}_{3},\ \check{g}\right]+8\Gamma^{\parallel} \check{g}_s \check{g}_{t} \sigma_y\;.
    \label{eq:AlgebraicUsadel}
\end{align}

Only the singlet and a single triplet component -- the one parallel to the external field -- are finite. This allows for rather compact expressions for the matrix current and torque to be derived from Eqs. (\ref{eq:matrixJ}-\ref{eq:matrixT}):
\begin{align}
    &\mathcal{\check{J}}^{S}_x = -\gamma \check{g}_t \left(\check{1} - \check{g}^2_t\right)+ \gamma \check{g}_s \check{g}^2_t \sigma_y\;,\\
    &\mathcal{\check{T}}^{S} = \frac{i \gamma}{2} \left(p^{S0}+\frac{eH_{0}d_{S}}{c}\right)\left[\check{\tau}_{3},\ \check{g}_{s} \right]\left(\check{1}-2\check{g}_{t}^{2}\right)\sigma_{y}\;.
\end{align}
We incorporated the fact that $ \mathcal{\check{J}}^S_z = 0$ due to the symmetry of our system, that is, $\alpha_{kz} = \gamma_{lz} = 0$, and $\mathcal{\check{J}}^S_y = 0$ because $\vec{p}^S \parallel \hat{\vec{x}}$. 

The self-consistency equation for $\check{\Delta}$ in terms of Matsubara frequencies reads 
\begin{align}\label{eq:DeltaSelfconsistency}
    \Delta = \frac{\lambda\pi T}{4}\sum_{\omega_n}\operatorname{tr}  \left\{\check{\tau}_{2}\check{g}(\omega_{n})\right\}.
\end{align}

In the following sections, we use Eqs.~(\ref{eq:pS_profile}) --(\ref{eq:DeltaSelfconsistency}) to calculate the response of junctions with interfacial SOC to the externally tunable magnetic field $\vec{H}_{0}$ and the current $\vec{I}_{0}$.  Firstly, we present  the G-L free energy functional with ISOC-related corrections. Then, we determine the critical temperature of the junctions using the G-L functional. Lastly, we estimate the superconducting diode efficiency and how it depends on the applied magnetic field and the intensities of the interfacial spin-charge coupling and the spin relaxation. 





\section{Ginzburg-Landau functional with ISOC induced corrections}
\label{sec:GL}

For temperatures close to the critical temperature, we can relate the action of the NLSM and the Usadel equation to the well-known G-L theory. With this, we determine the critical temperature of our junctions and make a connection to the results presented in \cite{devizorova2024interfacial}, where the G-L functional itself was determined based on symmetry arguments, and explain the similarities and differences between our results and theirs. We focus on how the G-L functional depends on $\gamma$, $\Gamma^{\parallel}$ and $d_S$, and set  the applied current to zero, $I_0 = 0$.

In this limit the Usadel equation reduces to a set of linear equations:
\begin{subequations}
    \begin{align}
        &-D(p^{S0})^{2}f_{s}+\frac{i\gamma}{d_{S}}\left(p^{S0}+\frac{eH_{0}d_{S}}{c}\right)f_{y}\operatorname{sgn}\omega\nonumber\\ &= 2|\omega_{n}| f_{s}+2ih\operatorname{sgn}\omega_{n}f_{y}+2i\Delta\;,\\
        &-D(p^{S0})^{2}f_{y}+\frac{i\gamma}{d_{S}}\left(p^{S0}+\frac{eH_{0}d_{S}}{c}\right)f_{s}\operatorname{sgn}\omega_{n}\nonumber\\ &= 2|\omega_{n}| f_{y}+2ih\operatorname{sgn}\omega_{n}f_{s}+\frac{8\Gamma^{\parallel}}{d_{S}}f_{y}\;.
    \end{align}
\end{subequations}
The solution to this linear system is
\begin{equation} \label{eq:Amplitudes}
\begin{aligned}
    &f_s =- \frac{i\Delta   \left( \vert \omega_n \vert + X + \frac{4 \Gamma^{\parallel} }{d_S}  \right)}{\left( \vert\omega_n \vert + X \right)\left( \vert\omega_n \vert + X  + \frac{4\Gamma^{\parallel} }{d_S} \right) +\left( \frac{\gamma p^{S}(0)}{2 d_S} - h\right)^2}\;, \\
    &f_y =  - \frac{ \Delta  \operatorname{sgn}{\omega_n}\left( g_L \mu_B H_0 - \frac{\gamma p^{S}(0)}{2d_S}  \right)}{\left( \vert\omega_n \vert + X \right)\left( \vert\omega_n \vert + X  + \frac{4\Gamma^{\parallel} }{d_S} \right) +\left( \frac{\gamma p^{S}(0)}{2 d_S} -h \right)^2}\;, \\
    &h = g_L \mu_B H_0\;,\quad X =  \frac{D p^2_{S0}}{2}\;,\quad p^{S}(0) = p^{S0} + \frac{e H_0 d_S}{c}\;. \\
\end{aligned}
\end{equation}

The details of the  derivation  of the G-L functional from the pair amplitudes up to the second order in $\Delta$ are shown in Appendix~\ref{subsec:GLderivationfromPairamplitudes}. The final result reads:

\begin{widetext}
\begin{equation}
    F_{GL}= \frac{T-T_{c0}}{T_{c0}} \vert\Delta \vert^2 +  \frac{( \pi p^{S0}\xi)^2}{4} \vert\Delta\vert^2 
    + 2M\left(\frac{4 \Gamma^{\parallel}}{\pi T_{c0} d_S}\right)  \left[ \frac{H_0}{H_{c2}}  \left(  \chi_{{H}} - \frac{\gamma }{2D}\right) - (p^{S0}\xi)  \frac{\gamma }{2 \pi T_{c0}\xi d_{S}}   \right]^2 \vert\Delta \vert^2. \label{eq:GL} 
\end{equation}    
\end{widetext}
Here, $\chi_H = \frac{g_L}{2 m D}$, $M(x)$ is defined as 
\begin{align}
    M(x) &= \sum_{n = 0}^{\infty}\frac{1}{(2n+1)^{2}(2n+1+x)} = \nonumber \\ 
    &= \frac{ \pi^2 x - 4 \psi^{(0)}\left(\frac{1+x}{2}\right) -4 \gamma_{E}   - 4 \log 4 }{8 x^2},
\end{align}
where $\gamma_E \approx 0.577$ is the Euler constant, $\psi^{(0)}(x)$ is the digamma function. This function monotonically decreases as a function of its argument. It 
has two important asymptotes, $M(0) = {7 \zeta(3)}/{8}$ with $\zeta$ 
being 
the Riemann zeta function, and $M(x\gg1) \approx {\pi^2}/{8 x}$. 
Moreover, we introduce $\chi_{H} = {g_{L}}/{2mD}$ for the simplicity of notation. 
We identify four different contributions in $F_{GL}$, Eq.~(\ref{eq:GL}):
\begin{equation}
    F_{GL} =  F^{(0)}_{GL} + F^{(K)}_{GL}  + F^{(H)}_{GL} + F^{(L)}_{GL}\;,
\end{equation}
where $F^{(0)}_{GL} = \frac{T - T_{c0}}{T_{c0}} \vert \Delta \vert^2$ is the usual BCS contribution that does not depend on the magnetic field or superconducting momentum and therefore represents the equilibrium free energy in the absence of a magnetic field. $F^{(K)}_{GL}$ represents the kinetic energy of the superconducting condensate that reads
\begin{equation}
\begin{aligned}\label{eq:FGLK}
    &F^{(K)}_{GL} = (p^{S0} \xi)^2  \left[\frac{\pi^2}{4} + \right.\left. 2M\left(\frac{4 \Gamma^{\parallel}}{\pi T_{c0} d_S}\right) \left(\frac{\gamma \xi}{Dd_S}\right)^2 \right]\vert \Delta \vert^2.
\end{aligned}
\end{equation}
The coefficient in square brackets  corresponds to the superconducting stiffness $W$ in Eq.~(\ref{eq:superconductingstiffnessMainText}), i.e. it describes how costly it is to impose a phase gradient. Meanwhile, $F^{(H)}_{GL}$ denotes the contribution proportional to the magnetic field, which in the absence of ISOC determines the critical magnetic field,
\begin{equation}\label{eq:Hpart}
    F^{(H)}_{GL} = 2\left(\frac{H_0}{H_{c2}}\right)^2  M\left(\frac{4 \Gamma^{\parallel}}{\pi T_{c0} d_S}\right) \left( \chi_{H}-\frac{\gamma }{2D}\right)^2 \vert \Delta \vert^2\;.
\end{equation}
Finally, the free energy contains a term which is  
first order in  both the superconducting momentum and the magnetic field, the so-called  Lifshitz invariant term \cite{mineev2008nonuniform,AgterbergInBook2012}. For a thin superconductor with ISOC, this term takes the form 
\begin{equation} \label{eq:Lifshitz_invariant}
    F^{(L)}_{GL} =  -2\left(\frac{H_0}{H_{c2}}p^{S0}\xi\right)M\left(\frac{4 \Gamma^{\parallel}}{\pi T_{c0} d_S}\right) \frac{\gamma \left( \chi_{{H}}-\frac{\gamma }{2D}\right) }{ \pi T_{c0}\xi d_{S}}\vert \Delta \vert^2.
\end{equation}

Both terms in Eq.~(\ref{eq:GL}) that depend on the superconducting momentum or magnetic field are squares, that is, they are positive. Thus the free energy is always larger than $F_{GL}^{(0)}$. From this we conclude that the critical temperature can not exceed $T_{c0}$. In other words,  magnetic fields can not increase the critical temperature  of the superconductor,as 
wrongly  predicted in Ref.~\cite{devizorova2024interfacial}.
Specifically, as shown in Appendix~\ref{subsec:Tc}, the critical temperature can be written as
\begin{equation} \label{eq:tc_00}
     \frac{T_{c}}{T_{c0}} = 1 - \beta  \left(\frac{H_{0}}{H_{c2}}\right)^2\;, 
\end{equation}
where $\beta$ is the dimensionless coefficient that obeys $ \beta  > 0$. The magnitude of $\beta$ depends on the spin-charge coupling both directly through Eq.~(\ref{eq:Hpart}) and through the dependence of the equilibrium phase gradient $p^{\text{an}}$ on $\gamma$. 
This equilibrium phase gradient is determined by minimization of the free energy in Eq.~(\ref{eq:GL}). Due to the presence of the Lifshitz invariant, Eq.~(\ref{eq:Lifshitz_invariant}), such a phase may be nonzero.
In this case, $p^{\text{an}}$ is called the anomalous phase gradient; it is  a precursor for the superconducting diode effect at lower temperatures.
It reads
\begin{align}\label{eq:ps0mgeneral}
    p^{\text{an}}\xi= \frac{8M \left(\frac{4\Gamma^{\parallel}}{\pi T_{c0}d_{S}}\right)\frac{\gamma}{D}\left(\chi_H-\frac{\gamma}{2D}\right)}{\pi^{2} d_S/\xi+4M\left(\frac{4\Gamma^{\parallel}}{\pi T_{c0}d_{S}}\right)\frac{\gamma^{2}}{ \pi T_{c0}D \xi d_S}}\frac{H_0}{H_{c2}}\;.
\end{align} 
This confirms that for a finite spin-galvanic coefficient,  $\gamma$ 
an external magnetic field $H_0$ induces a phase gradient.  This anomalous phase which is, according to Eq.~(\ref{eq:ps0mgeneral}),   suppressed in the limit of strong boundary spin relaxation. Notably, the Lifshitz invariant in Eq.~(\ref{eq:Lifshitz_invariant}) and consequently the anomalous superconducting momentum in Eq.~(\ref{eq:ps0mgeneral}) change sign when ${\gamma} = 2D \chi_H$. This value marks the boundary between the regimes in which the  triplets are predominantly generated by the exchange field or by vector potential in combination with  spin-charge coupling.

To compare our results with those of Ref.~\cite{devizorova2024interfacial}, we focus on the limit $d_S\xrightarrow{}0$. In that case, 
the coefficient 
in Eq.~(\ref{eq:Hpart}) vanishes, that is, $T_{c}$ may only depend on the external magnetic field through the generation of equilibrium gradients by $H_0$. On the other hand, the phase stiffness diverges, that is, imposing phase gradients becomes very costly. 

 Strikingly, the Lifshitz invariant converges to a finite value because $\lim_{d_S\xrightarrow{}0}\frac{\Gamma^{\parallel}}{\pi T_{c0}d_{S}}M\left(\frac{4\Gamma^{\parallel}}
{\pi T_{c0}d_{S}}\right) = \frac{\pi^{2}}{8}$, and therefore its coefficient is solely determined by the relative magnitude of the spin-galvanic and relaxation terms. 

The optimal phase gradient $p^{\text{an}}$ becomes
\begin{align}
    p^{\text{an}}\xi \approx \frac{1}{1+ \frac{8 D d_S\Gamma^{\parallel}}{\gamma^2}}\frac{Dd_{S}}{ \xi\gamma}\left(\chi_H -\frac{\gamma}{2D}\right)\frac{H_{0}}{H_{c2}}\;.
\end{align}
Thus, we find that, due to the increase of the superconducting stiffness, the optimal phase gradient is suppressed in very thin junctions.
The  parameter $\beta$ controlling the dependence of the critical temperature on the magnetic field in this limit reads
\begin{align}\label{eq:betathin}
    \beta = \left(\frac{\pi d_S D (\chi_{H}-\frac{\gamma}{2D})}{2\gamma\xi }\right)^{2}\;.
\end{align}
Thus, as $d_{S}\xrightarrow{}0$, $\beta$ converges to $0$, just like in materials without ISOC.

The difference between our result and Ref.~\cite{devizorova2024interfacial} is due to the suppression of the optimal phase gradient in our formalism, which can be traced back to the ISOC-dependent correction of the superconducting stiffness in Eq.~(\ref{eq:FGLK}).
The stiffness increases in the second order of $\gamma/D$. Usually, $\gamma/D$ is small and therefore this term is negligible if one considers junctions whose thickness is of the order of the coherence length $d_S/\xi \approx 1$. However, if the junction becomes thin, the phase stiffness increases as $\left(\frac{\gamma \xi}{D d_S}\right)^2$, so the phase gradients are distinctly harder to generate. This compensates for the free energy reduction by the Lifshitz invariant, by completing the square in Eq.~(\ref{eq:GL}).

\begin{figure}[hbtp]
    \centering
    \includegraphics[width = 0.99\linewidth]{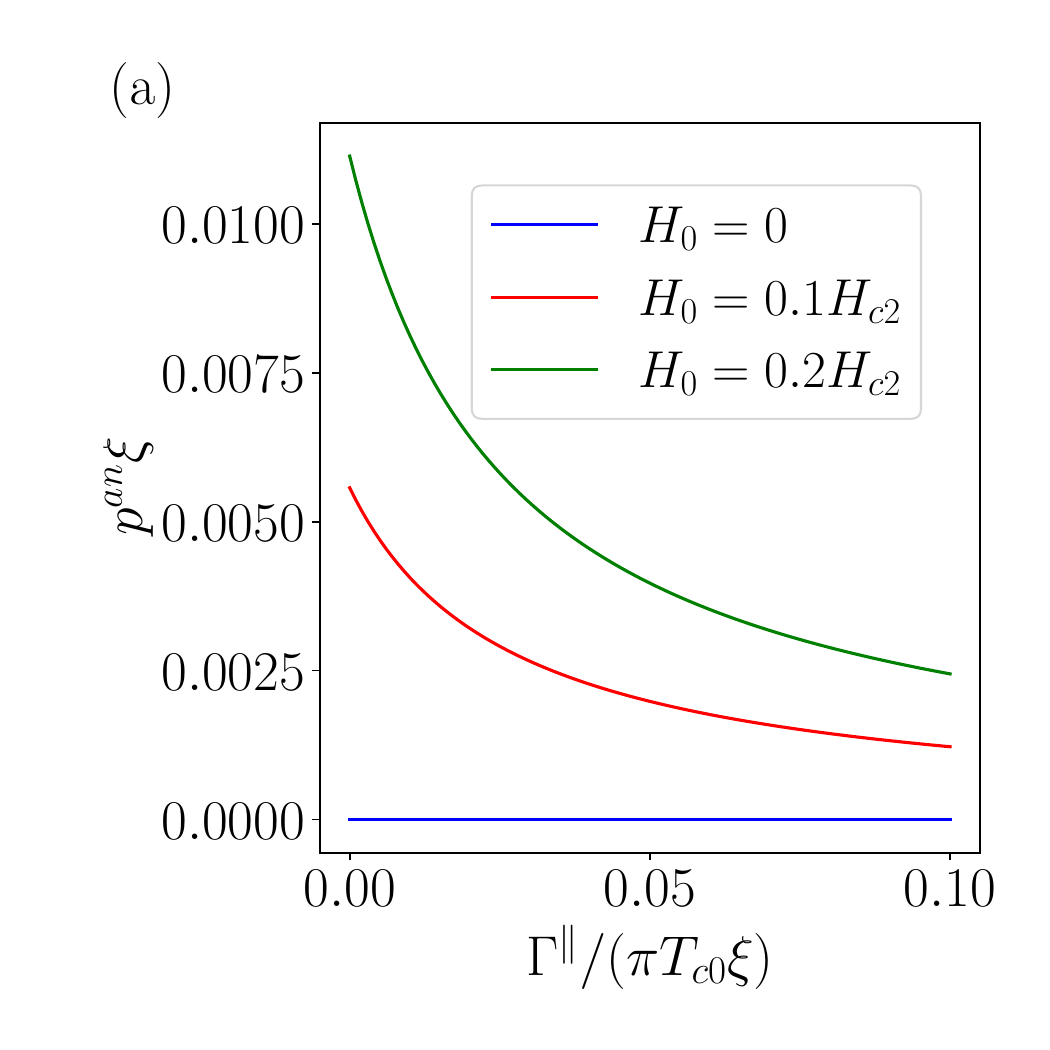}
    \includegraphics[width = 0.99\linewidth]{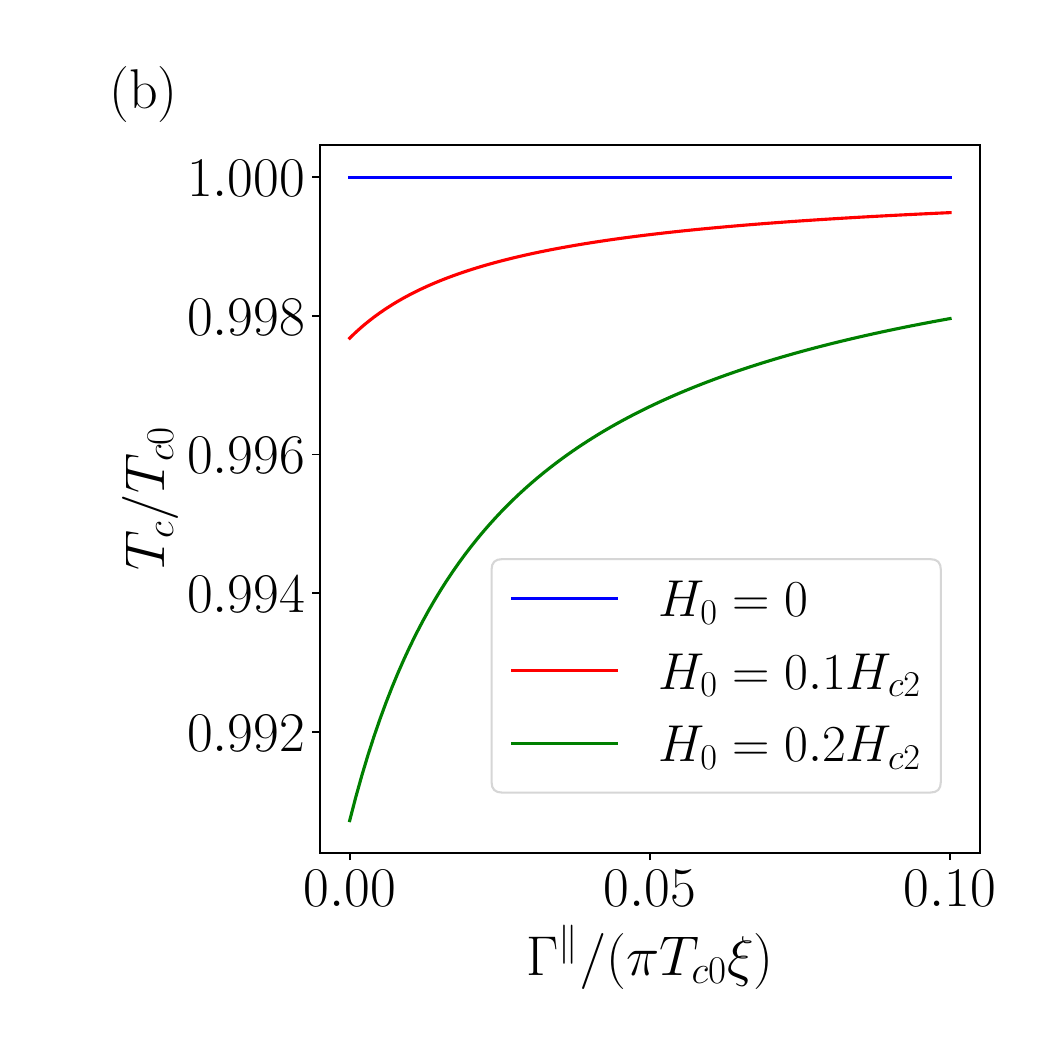}
    \caption{The effect of the spin-relaxation $\Gamma^\parallel$ on $p^{\text{an}}$ and $T_c$ given by Eqs.~(\ref{eq:tc_00}) and (\ref{eq:ps0mgeneral}). (a): With increasing spin-relaxation, the anomalous phase gradient decreases in magnitude. (b): As a result of this, the critical temperature is enhanced by spin relaxation. For strong spin relaxation, it converges to $T_{c0}$. The parameters of the superconducting film are $d_S = 0.1\xi$, $\frac{\gamma}{D} = 0.02$, $\chi_H = 0.35$.
    }
    \label{fig:Gamma_effect_Tc_p^{S}}
\end{figure}

Another interesting limit is 
the opposite limit, where the thickness is much shorter than the coherence length but larger than ${\gamma\xi}/{D}$. We find to first order in ${\gamma\xi}/{Dd_S}$
\begin{align}
    p^{\text{an}}\xi = 8M\left(\frac{4\Gamma^{\parallel}}{\pi T_{s0}d_S}\right)\frac{\gamma\xi}{\pi^{2}D d_S}\left(\chi_H-\frac{\gamma}{2D}\right)\frac{H_{0}}{H_{c2}}\;.
\end{align}
As expected, in this limit, even without spin relaxation ($M(0) =7\zeta(3)/8$), the optimal phase gradient is small  because the triplet correlations exist only near the boundary, while spin relaxation suppresses the optimal phase gradient, as illustrated in  Fig.~\ref{fig:Gamma_effect_Tc_p^{S}}(a). 
By substitution of this value in the free energy, we find that the coefficient $\beta$ governing the $H_{0}$ dependence of the critical temperature in this limit reduces to
\begin{align}\label{eq:betathick}
    \beta = 2 M\left(\frac{4\Gamma^{\parallel}}{\pi T_{s0}d_S}\right)\left( \chi_H -\frac{\gamma}{2D}\right)^{2}\;.
\end{align}
We note that even in this limit $\gamma$ has a strong influence on $\beta$, which is due to the vanishing of $\beta$ for small $d_{S}/\xi$ in the absence of ISOC. The size of $\beta$ is controlled by the interplay between the generation of triplets by the exchange field and the vector potential in combination with spin-charge coupling. Indeed, if these effects balance, no triplets are generated, and therefore the magnetic field does not affect the junction. If there is no perfect balance, triplets are generated and therefore magnetic fields suppress the pair amplitudes. As expected, since spin relaxation hampers the generation of triplets, it improves the stability of the superconducting phase against magnetic fields, thereby increasing the critical temperature in the presence of a magnetic field, as shown in Fig.~\ref{fig:Gamma_effect_Tc_p^{S}}(b).

\section{Superconducting diode effect}
\label{sec:Diode}

\begin{figure}[htbp]
    \centering
    \includegraphics[width=0.99\linewidth]{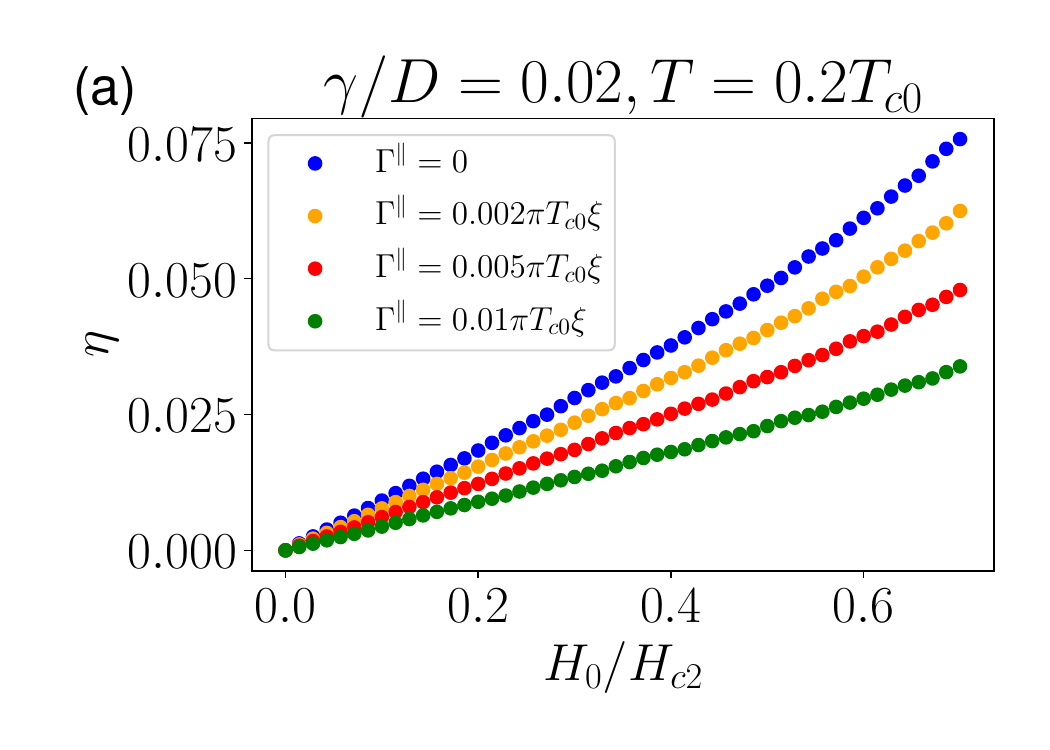}\\
    \includegraphics[width=0.99\linewidth]{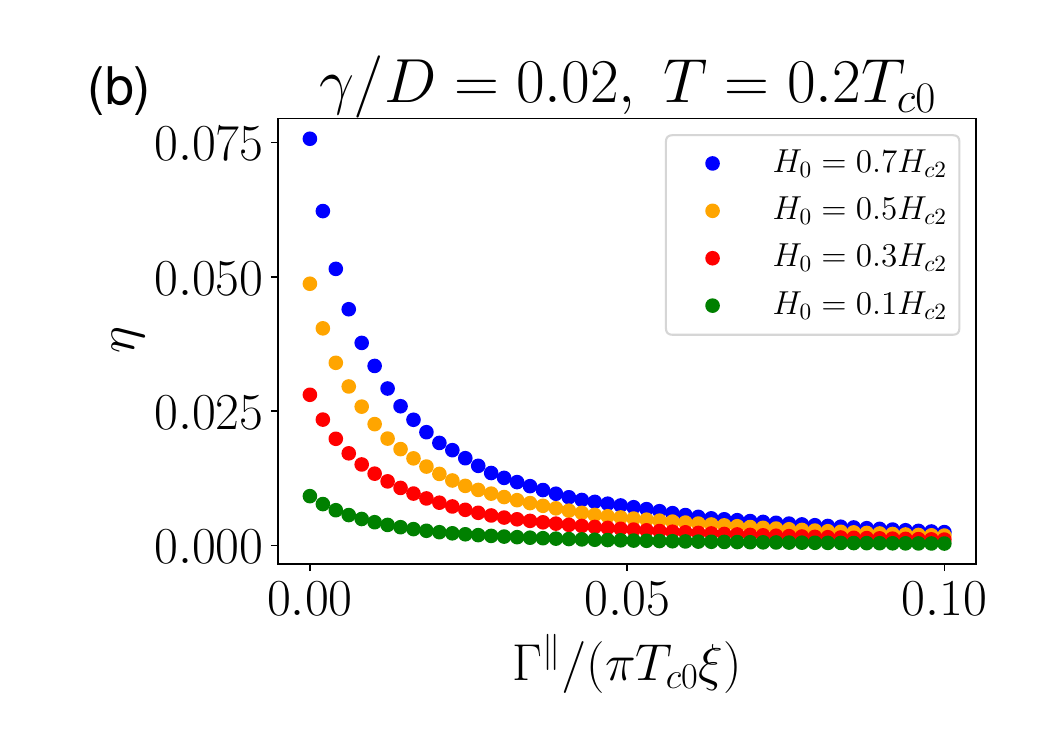}\\
    \includegraphics[width=0.99\linewidth]{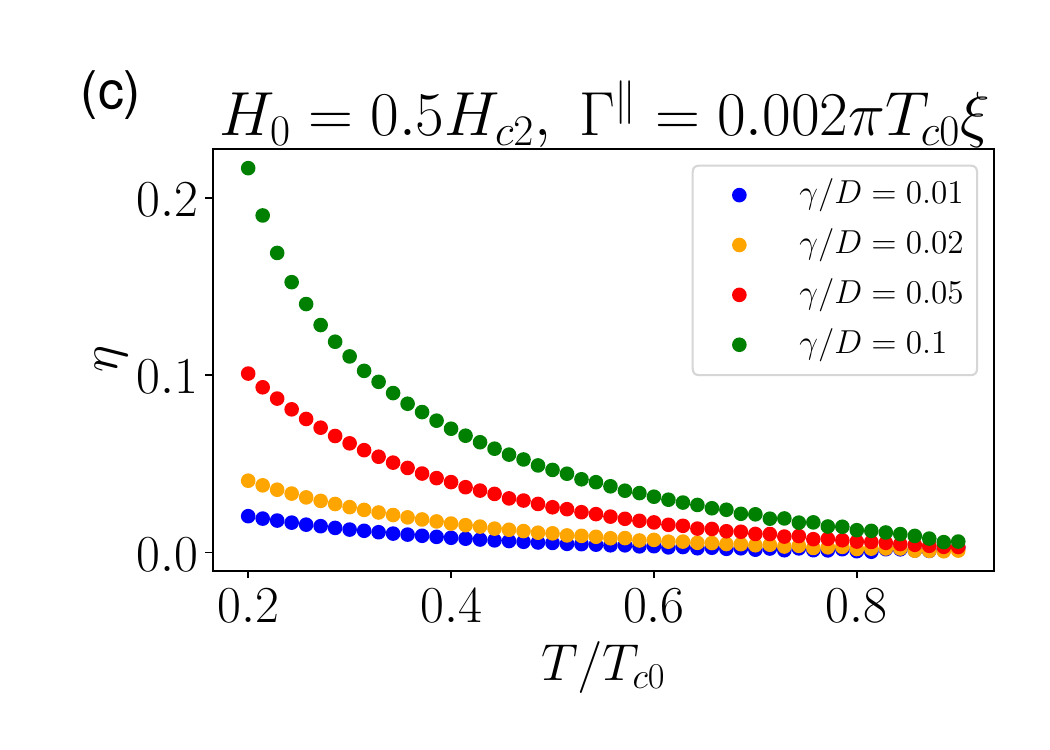}\\
    \caption{The dependence of the diode efficiency on different parameters of the system. 
    We choose $d_S = 0.2\xi$, $\chi_H \approx 0.35$. (a): The diode effect requires a magnetic field and increases monotonically with field strength in the range of parameters considered here. (b): Spin-relaxation suppresses the diode efficiency. 
    (c): The diode efficiency is largest for small temperatures and can reach $20\%$ for small spin-relaxation and a large spin-galvanic parameter.}
    \label{fig:diode_eta}
\end{figure}

From the appearance of the Lifshitz invariant in the G-L functional, we conclude that the system supports anomalous currents in the vicinity of the critical temperature. Therefore, for temperatures well below the critical one, we may expect the superconducting diode effect to appear. To investigate the dependence of this diode effect on the parameters that characterize the system, we numerically solve the nonlinear Usadel equation together with the self-consistency equation, specifically Eqs.~(\ref{eq:I0Eqmaintext}--\ref{eq:DeltaSelfconsistency}) using the numerical procedure described in Appendix~\ref{sec:Numerics_Details}. Subsequently, we find the diode efficiency as
\begin{align}
    \eta = \frac{I_{c+}-|I_{c-}|}{I_{c+}+|I_{c-}|}\;,
\end{align}
where $I_{c+}$ and $I_{c-}$ are the maximum supercurrents in the positive and negative directions. The results are shown in Figs.~\ref{fig:diode_eta}(a--c) and indicate the presence of a superconducting diode effect.

As discussed in the previous section, the spin-charge coefficient $\gamma$ is the key parameter that governs the intensity of ISOC-related effects.  
Experimental results \cite{sanz2020quantification,pham2021large,Groen23} on different insulator/normal metal systems indicate that $\sigma^{\text{sc}}$ can reach values of the order of
 $\sigma^{sc} \sim 10^5\ \ohm^{-1}\meter^{-1}$.
 For a typical superconductor such as Nb, the normal-state conductivity is of order
 $\sigma_D\sim 10^{7}\;\ohm^{-1}\meter^{-1}$ \cite{ashcroft1976solid}, this yields $\gamma/D=\sigma^{sc}/\sigma_D\sim 10^{-2}$.  


The dependence of $\eta$ on the applied magnetic field $H_0$ for different values of the spin relaxation coefficient is shown in Fig.~\ref{fig:diode_eta}(a).  We restrict ourselves to  external fields with  $g_L \mu_B H_0 < 0.5 \Delta$, to avoid  other phases,  such as the FFLO state \cite{fulde1964superconductivity,larkin1964nonuniform,buzdin2005review}. 
Since the superconducting diode effect requires time-reversal symmetry breaking, a nonzero diode efficiency necessitates the presence of a magnetic field. The effect gradually increases with increasing magnetic field, regardless of the strength of $\Gamma^{\parallel}$. However, as $\Gamma^{\parallel}$ increases, this growth in efficiency is significantly reduced, because spin relaxation suppresses the triplets and thereby weakens the effect of exchange fields.

The dependence of $\eta$ on spin relaxation is further illustrated in Fig.~\ref{fig:diode_eta}(b), where we observe that spin relaxation suppresses the diode efficiency. 
Existing experiments yield values for the so-called  spin mixing conductance $G^\parallel$ of heavy insulator/normal systems   
 typically of order $10^{13}\text{ -- }10^{14}\ \ohm^{-1}\meter^{-2}$
 \cite{Groen23}. 
Using a representative coherence length of 
$\xi\approx 10^{-8}\; \meter$ for Nb \cite{asada1969superconductivity,draskovic2013measuring,zaytseva2020upper}, we find that the dimensionless parameter ${\Gamma^{\parallel}}/{\pi T_{c0}\xi}= {G^\parallel \xi}/{4\sigma_D}$
is of order $10^{-2}\text{ -- }10^{-1}$.
For the value of  $\gamma/D = 0.02$ used above, and a small spin-relaxation coefficient $\Gamma^\parallel/(\pi T_{c0}\xi) \approx 10^{-2}$, we estimate from Fig.~\ref{fig:diode_eta}(b) that the 
diode efficiency 
can reach  $\approx 5\%$. 
This value can be improved by properly selecting material combinations and carefully engineering the interface to minimize $\Gamma^{\parallel}$ and maximize $\gamma$.

Finally, we show the dependence of the diode efficiency on the system temperature for different $\gamma$ coefficients in Fig.~\ref{fig:diode_eta}(c).  We see that the diode effect decreases as the system temperature increases. This happens because with increasing temperature, the higher harmonics decay faster than the first harmonic, so that the diode efficiency goes down even if the $\varphi_{0}$-effect remains. We find that for  $\gamma/D = 0.1$, even diode efficiencies of $\eta = 0.2$ are reachable for sufficiently low temperatures and relaxation rates.

\section{Conclusions}
\label{sec:conclusions}

We have presented a transport theory for hybrid structures of superconductors in contact with heavy-atom insulators that generate interfacial spin-orbit coupling (ISOC). The boundary conditions, Eqs.~(\ref{eq:Boundary_ISOC_main})-(\ref{eq:matrixT_Main_results}), derived from the effective action, allow us to describe transport properties in both the superconducting and normal states. In the vicinity of the critical temperature, from the obtained kinetic equations we have derived the Ginzburg-Landau functional for superconducting films with ISOC. We show that for thin junctions, the ISOC significantly enhances the phase gradient stiffness. This enhancement prevents the increase of $T_{c}$ by applying a magnetic field, in contrast to the suggestion of Ref.~\cite{devizorova2024interfacial} based on the Ginzburg-Landau theory with a naive inclusion of a surface Lifshitz invariant.  
We show how the combination of ISOC with an external magnetic field leads to anomalous currents and phase gradients. With the increase of the thickness of the junction, these effects are suppressed, because ISOC only affects the pair amplitudes close to the boundary, not throughout the junction. 

We also study the superconducting diode effect in these systems well below $T_{c}$.  We show that this effect is most clearly visible in thin films, in which the spin-galvanic effects dominate over the spin-relaxation effects. 
Thus, our results may motivate experimentalists in the search and fabrication  of  heavy atom insulator/superconductor interfaces with stronger spin-charge coupling parameter  ($\gamma$) but weaker spin relaxation ($\Gamma^{\parallel}$) to create superconducting diodes with conventional superconductors.

While in this paper we focused on isotropic boundaries, the developed theory can be used for non-isotropic interfaces. In such interfaces the spin-galvanic tensor $\gamma_{lk}$ and relaxation tensor $\Gamma_{ab}$, see Eq. (\ref{eq:SOC_action}),  will take different forms than those in Eqs. (\ref{eq:alphastructure},\ref{eq:gammastructure}).  In such a case, the dependence of the results on the magnitude of the spin-galvanic term will be qualitatively similar, but the dependence of the direction of the induced current on the direction of the magnetic field can be significantly altered, as was shown before for bulk spin-galvanic effects \cite{he2020magnetoelectric,kokkeler2024universal}.

\section{Acknowledgments}
\label{sec:grants}
We acknowledge financial support from Spanish MCIN/AEI/10.13039/501100011033 through Projects No. PID2023-148225NB-C31, No. PID2023-148225NB-C32 (SUNRISE) and No. TED2021-130292B-C42,  the Basque Government through grants No. IT-1591-22 and No. IT1453-22, and the European Union’s Horizon Europe research and innovation program under grant agreement No. 101130224 (JOSEPHINE). The computer program and its output used in this paper are available \cite{program2025}. 

\bibliographystyle{apsrev4-2}
\bibliography{refs.bib}
\clearpage
\appendix
\onecolumngrid

\section{Boundary Conditions in the normal state}\label{sec:NormalState}

In the normal state, superconducting correlations are absent, so $\Delta = 0$ and  the retarded, advanced and Keldysh components of the Green function $\check{g}$ in Eqs.~(\ref{eq:Usadel_main}) and (\ref{eq:Boundary_ISOC_main}) take the following form
\begin{align}
    \check{g}^{R}(t,t') &= -\check{g}^{A}(t,t') = \check{\tau}_{3}\delta(t-t')\;,\\
    \check{g}^{K}(t,t') &= 2\check{\tau}_{3}\check{F}(t,t')\;.
\end{align}
Therefore, we only need to solve for the distribution functions $F$. Moreover, because all the components commute with $\check{\tau}_{3}$, the resulting equations for $\check{F}(t,t')$ do not depend on the time difference coordinate $t-t'$, and therefore the equations for $t = t'$, which determine the physical observables, decouple from the equations for $t \neq t'$. 

The kinetic equation can in that case be directly written in terms of the excess charge and excess spin defined as
\begin{align}
    \delta n &= 2\nu_{0} \mu = -\frac{\pi\nu_{0}}{2} \operatorname{tr}\check{\tau}_{3}\check{F}(t,t)\;,\\
    \delta S_{a}& = 2\nu_{0} \mu_{a}^{s} =-\frac{\pi\nu_{0}}{2}\operatorname{tr}\sigma_{a}\check{F}(t,t)\;.
\end{align}

Using these definitions in the boundary condition, we find that  they can be written as
\begin{align}
    D \vec{n}\cdot\vec{\nabla} \delta n &= -\gamma \vec{n}\times\vec{\nabla}\cdot \delta\vec{S}\;,\\
    D\vec{n}\cdot\vec    {\nabla} \delta \vec{S}^{\parallel} &= -\gamma \vec{n}\times\vec{\nabla}\delta n + 8\Gamma^{\parallel} \delta \vec{S}^{\parallel} - 2\alpha (\vec{n}\times\vec{\nabla})\times\vec{n}\delta S^{\perp}\;,\\
    D\vec{n}\cdot\vec{\nabla}\delta S^{\perp}&=8\Gamma^{\perp}\delta S^{\perp}-2\alpha \vec{n}\cdot\left((\vec{n}\times\vec{\nabla})\times\delta\vec{S}\right)\;,
\end{align}
where for simplicity of notation we denoted $S^{\perp} = \vec{n}\cdot\vec{S}$ and $S^{\parallel} = \vec{S}-S^{\perp}\vec{n}$.
Comparing this with the result in \cite{borge2019boundary}, we find that we may identify $-\gamma/D$ with the spin-to charge conductivity \cite{sanz2020quantification} $\sigma^{sc}$, $\frac{-2\alpha}{D}$ with the spin-swapping coefficient $\kappa$ and $\Gamma^\parallel$ with the spin-loss coefficient $L/8$. Since we consider an interface with heavy atom insulators, in our model the conductance and spin conductance vanish. 

Since none of the terms constituting the boundary action contains Pauli matrices in Nambu space, the linearized boundary conditions for the pair amplitudes $f_{s}+\vec{f}_{t}\cdot\vec{\sigma}$ are obtained by substituting $\delta n\xrightarrow{}f_{s}$ and $\delta\vec{S}\xrightarrow{}\vec{f}_{t}$.

\section{Analytical derivations}
\label{sec:Approximations}

In this appendix, we present the technical details for some of the derivations in the main text. First, in Sec.~\ref{sec:Maxwell}, we show how the Maxwell equation can be solved in the thin film limit and derive Eq. (\ref{eq:I0Eqmaintext})
in the main text.  In Sec.~\ref{subsec:Self_Concictency}, we discuss the selfconsistency relation. In Sec.~\ref{subsec:Pair_amplitudes}, we solve the algebraic Usadel equation, Eq. (\ref{eq:AlgebraicUsadel}) for temperatures close to the critical temperature by linearizing it in the pair amplitudes. The obtained  pair amplitudes are subsequently used in Sec.~\ref{subsec:GLderivationfromPairamplitudes} to derive the G-L functional for systems with ISOC, which corresponds to Eq.~(\ref{eq:GL}) in the main text. Lastly, in Sec.~\ref{subsec:Tc}, we discuss how the critical temperature can be obtained from the G-L free energy, and find expressions for $T_{c}$ in various limits, thereby deriving Eqs.~(\ref{eq:betathin}) and (\ref{eq:betathick}) of the main text.

\subsection{Analytical solution of the Maxwell equations}
\label{sec:Maxwell}

In this subsection, we discuss how the Maxwell equation in Eq.~(\ref{eq:Maxwell})  can be solved in the thin film limit, and with this, we derive Eqs.~(\ref{eq:pS_profile}) and (\ref{eq:I0Eqmaintext}).

The Maxwell equation reads, according to Eq.~(\ref{eq:Maxwell}) of the main text:
\begin{align} \label{eq:Maxwell_Appendix}
    \nabla \times [\nabla\times\vec{A}] = \frac{4\pi}{c}\vec{j}_{s}\;,
\end{align}
where $\vec{j}_{s}$ is the supercurrent in the material that depends on the quasiclassical Green's function like discussed in the main text:
\begin{align}
    \vec{j}_{S} = W\vec{p}^{S}\;,
\end{align}
where $W = -\frac{e\pi \nu_{0}DT}{4}\sum_{\omega_n}\operatorname{tr}\left\{\check{\tau}_{3}\check{g}\left[\check{\tau}_{3},\ \check{g}\right]\right\}$ is the superfluid stiffness of the superconductor, and $\vec{p}^{S}$ is the superfluid momentum.
In the setup discussed within this paper, Eq.~\ref{eq:Maxwell_Appendix} is equipped with the boundary conditions presented in Eq.~(\ref{eq:Maxwell_BCs}) in the main text:
\begin{align}
    &\vec{m}\times \left(\nabla\times\vec{A}(d_{S}-0)-\vec{H}_{0}\right) = -\frac{2\pi}{c}\vec{I}_{0}\;,\\
    &\vec{n}\times\left(\nabla\times\vec{A}(+0)-\vec{H}_{0}\right) = \frac{4\pi}{c}\vec{J}^{S} -\frac{2\pi}{c}\vec{I}_{0}\;,
\end{align}
where $\vec{I}_{0}$ is the total current flowing through the material, which can be externally imposed, while $\vec{J}^{S}$ is the surface current, Eq.~(\ref{eq:matrixJ}), which in Matsubara space reads $J^{S} = -\frac{\pi \gamma}{8}\sum_{\omega_n}\operatorname{tr} \left\{ \check{\tau}_{3}\left\{\left[\sigma_{i},\ \check{g}\right],\ \sigma_{j}+\check{g}\sigma_{j}\check{g}\right\} \right\}$,
depends on the quasiclassical Green's function, but not on the vector potential or phase gradients.

The equation (\ref{eq:Maxwell_Appendix}) may be written in terms of the local magnetic field $\vec{H} = \nabla \times \vec{A}$ as
a first order differential equation
\begin{align}\label{eq:MaxwellH}
    \nabla \times \vec{H} = \frac{4\pi}{c}\vec{j}_{s}\;,
\end{align}
while its boundary conditions are
\begin{align}
    \vec{m} \times \vec{H} (d_{S}-0) &= \vec{m}\times \vec{H}_{0}-\frac{2\pi}{c}\vec{I}_{0}\;,\\
    \vec{n}\times \vec{H}(+0) &= \vec{n}\times \vec{H}_{0}+\frac{4\pi}{c}\vec{J}^{S} -\frac{2\pi}{c}\vec{I}_{0}\;.\label{eq:MaxwellBCH}
\end{align}
This is the differential equation of the first order with two boundary conditions; it only has a solution if
\begin{align}
    \vec{I}_{0} = \vec{J}^{S}+\int_{0}^{d_{S}}\vec{j}_{s}dz\;,
\end{align}
which reflects the fact that the bulk and the surface current should sum up to the total current.

As discussed in the main text, we consider $\vec{H}_{0} = H_{0}\hat{y}$, so that all currents flow in the $\hat{\vec{x}}$-direction. For each of them, like for $H_{0}$, we use plain notation to denote the magnitude of these vectors. Moreover, by choice of the gauge in Eq.~(\ref{eq:Phase_Rotation_Delta}), the local magnetic field and the bulk current only depend on the $z$ coordinate. With this, Eqs.~(\ref{eq:MaxwellH}--\ref{eq:MaxwellBCH}) can be simplified to
\begin{align}\label{eq:MaxwellHscalar}
    \partial_{z}H = -\frac{4\pi}{c}j_{S} = -\frac{4\pi}{c}Wp^{S}\;,
\end{align}
and
\begin{align}
    H(d_{S}) &= H_{0}-\frac{2\pi I_{0}}{c}\; ,\\
    H(0) &= H_{0}+\frac{2\pi I_{0}}{c}-\frac{4\pi}{c}J^{S}\;. 
\end{align}

In the thin limit, where $d_{S}\ll \xi$, we know that the quasiclassical Green's function is almost constant, and therefore, the superfluid stiffness $W$ is almost constant. Moreover, in the chosen gauge, there is no spatial dependence of the superconducting phase, so that $\vec{p}^{S}$ is directly related to $\vec{A}$ via $\vec{p}^{S} = -\frac{2e}{c}\vec{A}$, so that the magnetic field and superfluid momentum are related by
\begin{align}\label{eq:pSHrelation}
    H &= -\frac{c}{2e}\partial_{z}p^{S}\;.
\end{align}

Because $W$ is independent of position, Eqs.~(\ref{eq:MaxwellHscalar}--\ref{eq:pSHrelation}) can be solved analytically. The result is
\begin{align}
    p^{S}(z)& = -\frac{2e\lambda}{c}H_{0}\frac{\sinh{\frac{z-d_{S}/2}{\lambda}}}{\cosh{\frac{d_{S}}{2\lambda}}} + \frac{4 \pi e \lambda }{c^2}I_{0}\frac{\cosh{\frac{z-d_{S}/2}{\lambda}}}{\sinh{\frac{d_{S}}{2\lambda}}}-\frac{8 \pi e \lambda}{c^2 }J^{S}\frac{\cosh{\frac{z-d_{S}}{\lambda}}}{\sinh{\frac{d_{S}}{\lambda}}}\;,\\
    H(z) &= H_{0}\frac{\cosh{\frac{z-d_{S}/2}{\lambda}}}{\cosh{\frac{d_{S}}{2\lambda}}}-\frac{2\pi}{c}I_{0}\frac{\sinh{\frac{z-d_{S}/2}{\lambda}}}{\sinh{\frac{d_{S}}{2\lambda}}}+\frac{4\pi}{c}J^{S}\frac{\sinh{\frac{z-d_{S}}{\lambda}}}{\sinh{\frac{d_{S}}{\lambda}}}\;.
\end{align}

Here, for simplicity of notation we defined
\begin{align}
    \lambda = \frac{c}{\sqrt{8\pi e W}}\;,
\end{align}
which is the characteristic length scale over which the magnetic field decays, i.e. the temperature dependent London penetration depth. At zero temperature in a bulk superconductor, it reduces to the conventional expressions $W = \frac{\pi}{2} \sigma_{D}\Delta$ and thus $\lambda = \frac{c}{2\pi\sqrt{\sigma_{D}\Delta}} = \lambda_{L}$. With increasing temperatures, $W$ decreases and consequently $\lambda$ increases.
In dirty materials, the London penetration length is usually much larger than the coherence length \cite{schmidt1997physics}, since unlike the coherence length it is not directly suppressed by disorder. Therefore, $d_S\ll\xi\ll \lambda$, which means we may keep only terms up to linear order in $p^{S}$ and $H$:
\begin{align}
    H&\approx H_{0}-\frac{2\pi}{c }I_{0}\left(\frac{2z}{d_S}-1\right)+\frac{4\pi}{c} J^S\left(\frac{z}{d_S}-1\right)\;,\\
    p^{S}&\approx -\frac{2e}{c}H_{0}\left(z-\frac{d_{S}}{2} \right)+\frac{I_{0}-J^{S}}{Wd_{S}}\label{eq:PsExprI0}\;.
\end{align}
Defining the constant term in $p^{S}$ as $p^{S0}$, we find that
\begin{align}\label{eq:pSexprAppendix}
    p^{S}\approx-\frac{2e}{c}H_{0}\left(z-\frac{d_{S}}{2} \right)+p^{S0}\;,
\end{align}
with which we have derived Eq.~(\ref{eq:pS_profile}) in the main text. Moreover, comparing Eqs.~(\ref{eq:PsExprI0}) and (\ref{eq:pSexprAppendix}), we find that the homogeneous component of the phase and the total current in the junction are related via
\begin{align}
    I_{0} = Wd_{S}p^{S0}+J^{S}\;,
\end{align}
which is Eq.~(\ref{eq:I0Eqmaintext}) in the main text.

Lastly, we note that comparing the typical scales $H_{c2} = \frac{c\pi T_{c0}}{eD}$ and $I_{c}\sim \frac{\sigma_{D} \pi T_{c0}}{e} = 2e\nu_{0}D\pi T_{c0}$, we note that
\begin{align}
    \frac{cH_{c2}}{2\pi I_{c}}\sim \frac{1}{4\pi\nu_{0}}\left(\frac{c}{eD}\right)^{2} = 1.76\left(\frac{\lambda_{L}}{\xi}\right)^{2}\gg 1\;.
\end{align}
Thus, in the magnetic field, we may safely neglect the current-induced field compared to the contribution of the external field. This allows us to write the exchange term in terms of the external field $\vec{H}_{0}$ instead of the local field $\vec{H}(z)$ in Eq.~(\ref{eq:ExchangefromExternal}). In $p^{S}$ we should keep the current dependent terms because they are of lowest order in $\frac{d_S}{\lambda}$.

\subsection{The selfconsistency relation for \texorpdfstring{$\Delta$}{} }
\label{subsec:Self_Concictency}

The selfconsistency relation can be found by the variation of the action. Specifically, since the action vanishes for any causal configuration, we need to take the functional derivative of Eq.~(\ref{eq:Bulk_Action}) with respect to the quantum component of $\check{\Delta}^q$ \cite{kamenev2023field}. One finds the stationary point equation for $\Delta e^{i\varphi}$ in a form:
\begin{align} \label{eq:Self_Consistency}
    \Delta  e^{i\varphi}= \frac{ \pi \lambda}{8}\operatorname{tr} \Bigg( \check{\rho}_{1}\left(\check{\tau}_1 - i \check{\tau}_2\right)\check{g}(t,\ t) \Bigg)\;.
\end{align}
This relation corresponds to Eq.~(\ref{eq:Selfconsistency}) of the main text. In Matsubara domain, Eq.~(\ref{eq:Self_Consistency}) reads as
\begin{align}\label{eq:Selfconsistency1}
    \Delta e^{i\varphi} = \frac{ i\lambda \pi T}{4}\sum_{\omega_n}\operatorname{tr} \left\{(\check{\tau}_{1} - i \check{\tau}_2) \check{g} \right\}\;.
\end{align}
For the BCS uniform superconductor with the critical temperature $T_{c0}$, one can set $\varphi = 0$ everywhere, that gives the following equation:
\begin{align}
    \Delta = \lambda\pi T\sum_{\omega_n}\frac{\Delta}{\sqrt{\omega^{2}_n+\Delta^{2}}}\;,
\end{align}
which means
\begin{align}
    1 = \lambda \pi T\sum_{\omega_n}\frac{1}{\sqrt{\omega^{2}_n+\Delta^{2}}}\;.
\end{align}
As $T\xrightarrow{}T_{c0}$ we have 
\begin{align} \label{eq:Omegan_Sum_0}
    1 = \lambda\pi T_{c0}\sum_{ \omega_n}\frac{1}{ \vert \omega_n \vert} = \lambda\sum_{n>0}\frac{1}{n+\frac{1}{2}}\;.
\end{align}
This sum diverges if we leave $n$ unrestricted. However, in conventional superconductors, in which the pairing interaction is mediated by phonons, there is only attractive interaction if $\omega_{n}<\omega_{D}$, where $\omega_{D}$ is the Debye frequency. Consequently, $n$ should be less than $\frac{\omega_{D}}{2\pi T_{c0}}$, and
\begin{align}
    \frac{1}{\lambda} \approx \log \left(\frac{\omega_{D}}{\pi T_{c0}} \right)\;.
\end{align}
Using this, we may write Eq.~(\ref{eq:Selfconsistency1}) as
\begin{align}\label{eq:Omegan_Sum_1}
    \Delta e^{i\varphi} \log \left(\frac{\omega_{D}}{\pi T_{c0}}\right) = \frac{i \pi T}{4}\sum_{\omega_n }  \operatorname{tr}\left\{ (\check{\tau}_1 - i \check{\tau}_2)\check{g} \right\}\;.
\end{align}
With the help of the same trick as in Eq.~(\ref{eq:Omegan_Sum_0}) to restrict the summation over $\omega_n$ up to $\omega_D$ for any temperature $T$, we write:
\begin{align} \label{eq:Omegan_Sum_2}
    \pi T \sum_{\omega_n}\frac{\Delta e^{i\varphi}}{\vert \omega_n \vert } = \Delta e^{i\varphi} \log \left(\frac{\omega_{D}}{\pi T}\right)\;.  
\end{align}
Subtracting Eq.~(\ref{eq:Omegan_Sum_1}) from Eq.~(\ref{eq:Omegan_Sum_2}), we find
\begin{align} \label{eq:Selfconsistency2}
    \Delta e^{i\varphi} \log\frac{T_{c0}}{T} =  \pi T \sum_{\omega_n} \left(\frac{\Delta e^{i \varphi}}{ \vert \omega_n \vert} - \frac{i }{4} \operatorname{tr}\left\{(\check{\tau}_1 - i \check{\tau}_2) \check{g}\right\} \right).
\end{align}

\subsection{Pair amplitudes in a thin superconducting film with ISOC}
\label{subsec:Pair_amplitudes}

In the vicinity of the critical temperature, $T_{c0} - T \ll T_{c0}$, we linearize Eq.~(\ref{eq:AlgebraicUsadel}) near the normal metal solution, that is, we write
\begin{align}
    \check{g} \approx \operatorname{sgn}\omega_{n} \check{\tau}_{3}+\begin{pmatrix}
        0&f_{s}+f_{y}\sigma_{y}\\
        -f_{s}-f_{y}\sigma_{y} & 0 \\
    \end{pmatrix}\;,
\end{align}
and substitute this into Eq.~(\ref{eq:AlgebraicUsadel}). Keeping only the terms of the first order in the pair amplitudes in the resulting equation, we obtain
\begin{subequations}
    \begin{align}
        -D(p^{S0})^{2}f_{s}+\frac{i\gamma}{d_{S}}\left(p^{S0}+\frac{eH_{0}d_{S}}{c}\right)f_{y}\operatorname{sgn}\omega = 2|\omega_{n}| f_{s}+2ih\operatorname{sgn}\omega_{n}f_{y}+2i\Delta\;,\\
        -D(p^{S0})^{2}f_{y}+\frac{i\gamma}{d_{S}}\left(p^{S0}+\frac{eH_{0}d_{S}}{c}\right)f_{s}\operatorname{sgn}\omega_{n} = 2|\omega_{n}| f_{y}+2ih\operatorname{sgn}\omega_{n}f_{s}+\frac{8\Gamma^{\parallel}}{d_{S}}f_{y}\;.
    \end{align}
\end{subequations}
The solution to this linear algebraic equation is is
\begin{equation} \label{eq:Amplitudes_Appendix}
\begin{aligned}
    &f_s =- \frac{i\Delta   \left( \vert \omega_n \vert + X + \frac{4 \Gamma^{\parallel} }{d_S}  \right)}{\left( \vert\omega_n \vert + X \right)\left( \vert\omega_n \vert + X  + \frac{4\Gamma^{\parallel} }{d_S} \right) +\left( \frac{\gamma p^{S}(0)}{2 d_S} - g_L \mu_B H_0 \right)^2}\;, \\
    &f_y =  - \frac{ \Delta  \operatorname{sgn}{\omega_n}\left( g_L \mu_B H_0 - \frac{\gamma p^{S}(0)}{2d_S}  \right)}{\left( \vert\omega_n \vert + X \right)\left( \vert\omega_n \vert + X  + \frac{4\Gamma^{\parallel} }{d_S} \right) +\left( \frac{\gamma p^{S}(0)}{2 d_S} - g_L \mu_B H_0 \right)^2}\;, \\
    & X =  \frac{D p^2_{S0}}{2}\;,\qquad p^{S}(0) = p^{S0} + \frac{e H_0 d_S}{c}\;. \\
\end{aligned}
\end{equation}

\subsection{Derivation of the Ginzburg-Landau functional}\label{subsec:GLderivationfromPairamplitudes}

The pair amplitude $f_s$ derived in Appendix~\ref{subsec:Pair_amplitudes} can be expanded up to the second order in $\frac{H_0}{H_{c2}}$ and $p^{S0}\xi$ as follows
\begin{align}
    f_{s} &= -i \vert \Delta \vert e^{i\varphi} \left\{\frac{1}{\vert \omega_n\vert} - \left(\frac{H_0}{H_{c2}} \right)^2 \frac{1}{\omega^2_n}\frac{\left(g_L \mu_B H_{c2}-\frac{\gamma \pi T_{c0}}{2D}\right)^{2}}{  \vert \omega_n \vert+ 4\Gamma^{\parallel}/d_S} - (p^{S0}\xi)^{2} \frac{\pi T_{c0}}{\omega^2_n}\left[1+ \frac{\gamma^{2} }{2 D d_{S}^{2}} \frac{1}{ \vert \omega_n \vert + 4\Gamma^{\parallel}/d_S}\right] + \right. \nonumber\\& \left.+  \left(p^{S0}\xi\frac{H_0}{H_{c2}} \right) \frac{ \gamma }{\xi d_S}  \frac{1}{\omega^2_n} \frac{\left( g_L \mu_B H_{c2} -\frac{\gamma \pi T_{c0}}{2D}\right)}{ \vert \omega_n \vert + 4\Gamma^{\parallel}/d_S }\right\}\;.
\end{align}
We restore the global phase  $e^{i\varphi}$ in  the order parameter,  $\Delta =  \vert \Delta \vert e^{i\varphi}$.  By  inserting  this expansion into the selfconsistency equation, Eq.~(\ref{eq:Selfconsistency2}) we obtain:
\begin{align}
    0 &= -\vert \Delta \vert e^{i\varphi} \log\frac{T_{c0}}{T}+\pi T \sum_{\omega_{n}}\left[\frac{ \vert \Delta \vert e^{i\varphi}}{\vert \omega_{n} \vert }-if_{s}\right]\approx \vert\Delta \vert e^{i\varphi} \frac{T-T_{c0}}{T_{c0}}+\pi T \sum_{\omega_{n}}\left[\frac{ \vert \Delta \vert  e^{i\varphi}}{\vert\omega_{n}\vert}-if_{s}\right]\;,
\end{align}
where we expanded the logarithm up to the first order in $T-T_{c0}$, consistent with the G-L limit.
We note that the resulting selfconsistency equation has to be the saddle point of the corresponding Ginzburg-Landau functional with respect to $\Delta^\dagger = \vert\Delta \vert e^{-i\varphi}$. Therefore, we obtain the G-L functional up to quadratic terms in $\Delta$:
\begin{equation} \label{eq:G-L_appendix}
\begin{aligned}
    &F_{GL}   = \frac{T - T_{c0}}{T_{c0}} \vert \Delta \vert^2 +  2 \vert\Delta \vert^2 \pi T_{c0} \sum_{\omega_n > 0} \left[\frac{1}{\omega_n} - \frac{i f_s}{\Delta}\right] = \\
    &= \frac{T-T_{c0}}{T_{c0}} \vert\Delta \vert^2 +  \frac{( \pi p^{S0}\xi)^2}{4} \vert\Delta\vert^2 + 2M\left(\frac{4 \Gamma^{\parallel}}{\pi T_{c0} d_S}\right)  \left[ \frac{H_0}{H_{c2}}  \left( \chi_H -\frac{\gamma }{2D}\right) - (p^{S0}\xi)  \frac{\gamma }{2 \pi T_{c0}\xi d_{S}}   \right]^2 \vert\Delta \vert^2\;.
\end{aligned}
\end{equation}
Here $M(x)$ is the dimensionless function that is defined as follows,
\begin{equation}
    M(x) = \sum_{n = 0}^{\infty}\frac{1}{(2n+1)^{2}(2n+1+x)} = \frac{ \pi^2 x - 4 \psi^{(0)}\left(\frac{1+x}{2}\right) -4 \gamma_{E}   - 4 \log 4 }{8 x^2}\;,
\end{equation}
where $\gamma_E \approx 0.577$ is the Euler constant, $\psi^{(0)}(x)$ is the digamma function. This function  monotonically decreases; it has two important asymptotics, $M(0) = \frac{7 \zeta(3)}{8}$ with $\zeta(3) \approx 1.202$ being the corresponding value of the Riemann zeta function, and $M(x\gg1) \approx \frac{\pi^2}{8 x}$. With this, we have derived Eq.~(\ref{eq:GL}) in the main text.

\subsection{The shift of critical temperature}
\label{subsec:Tc}

We may now infer the change in the critical temperature caused by ISOC-induced corrections using the definition of the critical temperature as the highest temperature at which superconductivity appears. The quadratic terms of the Ginzburg-Landau functional always take the form
\begin{align}\label{eq:FGLTc}
    F_{GL}\left[\Delta\right] = \frac{T-T_{c}}{T_{c}}\vert\Delta\vert^2\;.
\end{align}
Comparing Eqs.~(\ref{eq:G-L_appendix}) and (\ref{eq:FGLTc}), we infer that the critical temperature is given by
\begin{align} \label{eq:Tc_pS}
    \frac{T_{c}(p^{S0},\ H)}{T_{c0}} = 1-\frac{(\pi p^{S0}\xi)^{2}}{4}-2M\left(\frac{4\Gamma^{\parallel}}{\pi T_{c0} d_{S}}\right)\left[\frac{H_0}{H_{c2}}  \left( \chi_H -\frac{\gamma }{2D}\right) - (p^{S0}\xi)  \frac{\gamma }{2 \pi T_{c0}\xi d_{S}}\right]^{2}\;.
\end{align}

In equilibrium, the system acquires the phase gradient that minimizes the free energy, that is, that maximizes the critical temperature. It can thus be written as
\begin{align}
    \frac{T_{c}(H)}{T_{c0}} = \operatorname{max}_{p^{S0}} 
    \left\{1-\frac{(\pi p^{S0}\xi)^{2}}{4}-2M\left(\frac{4\Gamma^{\parallel}}{\pi T_{c0} d_{S}}\right)\left[\frac{H_0}{H_{c2}}  \left( \chi_H -\frac{\gamma }{2D}\right) - (p^{S0}\xi)  \frac{\gamma }{2 \pi T_{c0}\xi d_{S}}\right]^{2} \right\} \;.
\end{align}
The maximum of the critical temperature, which corresponds to the minimum of the free energy, is reached at  the optimal momentum $p^{\text{an}}$ that is given by
\begin{equation}
    p^{\text{an}} \xi = \frac{8M \left(\frac{4\Gamma^{\parallel}}{\pi T_{c0}d_{S}}\right)\frac{\gamma}{D}\left(\chi_H-\frac{\gamma}{2D}\right)}{\pi^{2} d_S/\xi+4M\left(\frac{4\Gamma^{\parallel}}{\pi T_{c0}d_{S}}\right)\frac{\gamma^{2}}{ \pi T_{c0}D \xi d_S}}\frac{H_0}{H_{c2}}\;,
\end{equation}
where we introduce $\chi_H = \frac{g_L \mu_B H_{c2}}{\pi T_{c0}} = \frac{g_L}{2 m D}$. At $\Gamma^\parallel = 0$, $p^{\text{an}}$ may be written as
\begin{equation} \label{eq:p^{S}0m}
\begin{aligned}
   &p^{\text{an}}\xi = \frac{7 \zeta(3)\frac{\gamma}{D}\left(\chi_H-\frac{\gamma}{2D}\right)}{\pi^{2} d_S/\xi+\frac{ 7\zeta(3) \gamma^{2} \xi}{D^2 d_S}}\frac{H_0}{H_{c2}} \approx  \frac{H_0}{H_{c2}}\left(\chi_H-\frac{\gamma}{2D}\right) \times\begin{cases}
        \frac{D d_S}{\gamma \xi}\;,\qquad d_S \ll \frac{\gamma \xi}{D}\;,\\
        \frac{7 \zeta(3) \gamma \xi}{ \pi^2 D d_S}\;, \qquad d_S \gg \frac{\gamma \xi}{D}\;. 
   \end{cases}
\end{aligned}
\end{equation}
The presence of the maximum in $T_c(p^{\text{an}})$ at a nonzero $p^{\text{an}}$ suggests the existence of a helical phase, which in turn implies non-reciprocal transport. It is worth noting that $p^{\text{an}}$ is suppressed in both the limits of very small and very large thicknesses, indicating a non-monotonic behavior. Consequently, the helical phase is most prominent when the thickness $d_S$ satisfies $d_{S} \sim \frac{\gamma \xi}{D}$.

For $H_0 = 0$, we have $p^{\text{an}} = 0$ and consequently $T_c = T_{c0}$, as expected. For nonzero $H_0$, one may write the critical temperature as
\begin{equation}
     \frac{T_{c}}{T_{c0}} = 1 - \beta \frac{H^2}{H^2_{c2}}\;, 
\end{equation}
where $\beta$ is the dimensionless coefficient that obeys $\beta > 0$. Specifically, in the very thin  limit $d_{S}\ll \frac{\gamma \xi}{D}$, we have by direct substitution of the small $d_{S}$ limit of Eq.~(\ref{eq:p^{S}0m}) into Eq.~(\ref{eq:Tc_pS}) and keeping terms only to lowest order in $d_{S}$:
\begin{align}
    \beta = \left(\frac{\pi (\chi_H-\frac{\gamma}{2D}) Dd_{S}}{2\gamma\xi} \right)^2\;.
\end{align}
This corresponds to Eq.~(\ref{eq:betathin}) in the main text.

On the other hand, if we take $p^{\text{an}}$ corresponding to the case of $\frac{\gamma \xi}{D} \ll d_S$, insert it into Eq.~(\ref{eq:Tc_pS}) and expand the result up to the second order in $\gamma/D$, then we find,
\begin{equation} \label{eq:Buzdin}
\begin{aligned}
    &\beta = \frac{7\zeta(3)}{4}\left( \chi_{H}-\frac{\gamma}{2D}\right)^{2}-\left(\frac{7\zeta(3)\gamma\xi \chi_H }{2\pi D d_{S}}\right)^{2}\;,
\end{aligned}
\end{equation}
which in the limit of very large $d_{S}$ reduces to Eq.~(\ref{eq:betathick}) in the main text.
Interestingly, if we subsequently wrongly put the opposite limit $d_S \ll \frac{\gamma \xi}{D}$ in Eq.~(\ref{eq:Buzdin}) instead of the correct limit one, $d_S \gg  \frac{\gamma \xi}{D}$, then we have the same dependence on $T_c(H_0)$ as reported in \cite{devizorova2024interfacial} using the G-L functional, where the positive contribution to $T_c$ dominates.

\clearpage
\section{Numerical procedures}
\label{sec:Numerics_Details}

In this appendix, we discuss the numerical code used to generate Fig.~\ref{fig:diode_eta} based on Eqs.~(\ref{eq:AlgebraicUsadel}) and (\ref{eq:Selfconsistency}) of the main text. 

\subsection{Dimensionless variables}
For numerical procedures, it is usually convenient to reduce the number of free parameters by using dimensionless parameters. The spin-galvanic term at the boundary has the same dimensions as the diffusion constant, and therefore, we define
\begin{align}
    \theta_{I} = \frac{\gamma}{D} = -2\frac{\sigma^{sc}}{\sigma_{D}}\;,
\end{align}
where $\sigma^{sc}$ is the spin-charge conductivity \cite{borge2019boundary} and $\sigma_{D}$ is the normal state conductivity of the superconductor.

For the other quantities, we use typical scales of the superconductor. 
For lengths and momenta, we use the superconducting coherence length $\xi = \sqrt{\frac{D}{2\pi T_{c0}}}$:
\begin{align}
    \tilde{d}_S&=d_S/\xi\;,\qquad
    \tilde{p}^{S}=p^{S}\xi\;.
\end{align}

For energies and temperatures, we use the BCS critical temperature $T_{c0}$:
\begin{align}
    \tilde{T} = \frac{T}{T_{c0}}\;,\qquad\tilde{\Delta} = \frac{\Delta}{\pi T_{c0}}\;,\qquad\tilde{\omega}_{n} = \frac{\omega_{n}}{\pi T_{c0}} = \tilde{T}(2n+1)\;.
\end{align}
For magnetic fields, we use the critical field $H_{c2}$ of the superconductor:
\begin{align}
    \tilde{H}(z) &= \frac{H(z)}{H_{c2}}\;,\qquad
    \tilde{H}_{0} = \frac{H_{0}}{H_{c2}}\;,
\end{align}
while for current densities and currents, we use the scales $j_{s0} = \frac{\sigma_{D}\pi T_{c0}}{\xi e}$ and $j_{s0}\xi$, related to the critical current density:
\begin{align}
    \tilde{j}_{S}(z) &= \frac{j_{S}(z)}{j_{s0}}\;,\qquad
    \tilde{W} = \frac{W}{j_{s0}\xi}\;,\qquad \tilde{J}^{S} = \frac{J^{S}}{\xi j_{s0}}\;,\qquad \tilde{I}_{0} = \frac{I_{0}}{\xi j_{s0}}\;.
\end{align}
The boundary spin-relaxation, which has units of velocity, is normalized via
\begin{align}
    \tilde{\Gamma}^\parallel = \frac{\Gamma^{\parallel}}{\pi T_{c0}\xi}\;.
\end{align}
With this, the algebraic Usadel equation (\ref{eq:AlgebraicUsadel}) reads as
\begin{align} \label{eq:AlgebraicUsadel_d1}
    - \frac{(\tilde{p}^{S0})^{2}}{2}\left[\check{\tau}_{3},\ \check{g}\left[\check{\tau}_{3},\ \check{g}\right]\right] -\frac{i \tilde{p}_{S}(0)}{\tilde{d}_{S}}\left[\check{\tau}_{3},\ \check{\tilde{J}}^S\right]+\mathcal{\check{\tilde{T}}}^S = \left[\tilde{\omega}_{n}\check{\tau}_{3}+\tilde{\Delta}\check{\tau}_{2}+i\chi_{H}\tilde{H}_{0} \sigma_y\check{\tau}_{3},\ \check{g}\right]+\frac{8\tilde{\Gamma}^\parallel}{\tilde{d}_{S}}\check{g}_{s}\check{g}_{t}\sigma_{y}\;,
\end{align}
where for simplicity of notation we introduced

\begin{align} \label{eq:Dimensionless_Parameters}
    &\chi_{H} = \frac{g_L \mu_B H_{c2} }{\pi T_{c0}} = \frac{g_L}{2 m D}\;.
\end{align}

and the dimensionless surface current and torque read
\begin{equation}\label{eq:AlgebraicUsadel_d2}
\begin{aligned}
    \check{\tilde{J}}^S &= -\theta_{I}\check{g}_{t}(\check{1}-\check{g}_{t}^{2})+\theta_{I}\check{g}_{s}\check{g}_{t}^{2}\sigma_{y}\;,\\
    \mathcal{\check{\tilde{T}}}^S &= \frac{i\theta_{I} \tilde{p}^{S}(0)}{\tilde{d}_{S}}\left[\check{\tau}_{3},\ \check{g}_{s}\right]\left(\check{1}-2\check{g}_{t}^{2}\right)\sigma_{y}\;.
\end{aligned}
\end{equation}

The selfconsistency relation, Eq.~(\ref{eq:DeltaSelfconsistency}), within dimensionless units reads
\begin{align}\label{eq:Selfconsistencydimensionless}
    -\tilde{\Delta}  \log \tilde{T} = \sum_{n}\left[ \frac{\tilde{\Delta}}{\vert 2n+1 \vert }-\frac{i\tilde{T}}{4}\operatorname{tr}\{(\check{\tau}_{1}-i\check{\tau}_{2})\check{g}\} \right] \;.
\end{align}
Meanwhile, Eq.~(\ref{eq:I0Eqmaintext}) that relates the total current to the surface current and the bulk current reads
\begin{align}\label{eq:I0dimensionless}
    \tilde{I} = \tilde{J}^{S}+\tilde{W}\tilde{d}_S\tilde{p}^{S}\;.
\end{align}

\subsection{Parameterization}
In order to solve Eqs.~\ref{eq:AlgebraicUsadel_d1}, \ref{eq:AlgebraicUsadel_d2}, and \ref{eq:Selfconsistencydimensionless}, we employ the angle parameterization:
\begin{equation} \label{eq:Green_Function_Paramtrization}
\begin{aligned}
    &\check{g} = \begin{pmatrix}
        \hat{g} & \hat{f} \\
        -\hat{\tilde{f}} & -\hat{\tilde{g}}
    \end{pmatrix} = \cosh\zeta \begin{pmatrix}
		\cosh\theta & \sinh\theta\\
		-\sinh\theta & -\cosh\theta\end{pmatrix} +  \sigma_y \sinh \zeta \begin{pmatrix}\sinh\theta & \cosh\theta\\
		-\cosh\theta & -\sinh\theta\end{pmatrix}.
\end{aligned}
\end{equation} 
The gauge transformation, Eq.~(\ref{eq:Phase_Rotation_Delta}), makes $\check{\Delta}$ real, so $\hat{\tilde{f}} = \hat{f}$, $\hat{\tilde{g}} = \hat{g}$, that is, $\zeta$ is real and $\theta$ is imaginary. In terms of this parameterization, the algebraic Usadel equation (\ref{eq:AlgebraicUsadel_d1}) reads:
\begin{equation} \label{eq:Algebraicparameterized}
\begin{aligned}
&\frac{i \theta_I}{\tilde{d}_S}  \tilde{p}^S(0) \sinh \zeta \cosh^2 \zeta  \cosh \theta -\frac{{ \left( \tilde{p}^{S0} \right)^2}}{2} \cosh 2 \zeta \sinh 2 \theta -
	\tilde{\omega}_n  \cosh \zeta  \sinh \theta  -  i \cosh \theta  \left[\tilde{\Delta}  \cosh \zeta  + \chi_{H} \tilde{H}_0 \sinh \zeta  \right] = 0\;, \\
	&\frac{i \theta_I }{4\tilde{d}_S}  \tilde{p}^S(0) \left[\cosh \zeta+3 \cosh 3\zeta  \right] \sinh \theta  - \frac{2\tilde{\Gamma}_i}{\tilde{d}_S} \sinh 2\zeta -
 \\ 
    &\qquad\qquad-\frac{{ \left( \tilde{p}^{S0} \right)^2}}{2} \sinh 2 \zeta  \cosh 2 \theta  - \tilde{\omega}_n  \sinh \zeta  \cosh \theta  - i \sinh \theta \left[\tilde{\Delta}  \sinh \zeta + \chi_{H} \tilde{H}_0 \cosh \zeta \right] = 0\;.    
\end{aligned}
\end{equation}
These equations need to be solved for each $\tilde{\omega}_{n}$. We denote their solutions by $\theta_{n},$ $\zeta_{n}$. The dimensionless selfconsistency equation, Eq.~(\ref{eq:Selfconsistencydimensionless}), in this parameterization reads
\begin{align}
    -\tilde{\Delta} \log\tilde{T} = 2 \operatorname{Re}\sum_{\omega_n > 0} \left[\frac{\tilde{\Delta}}{2n+1} - i\tilde{T} \cosh \zeta_{n} \sinh\theta_{n} \right]\;.
\end{align}

After the equation for selfconsistency is solved, the dimensionless superfluid stiffness and surface currents  are calculated as
\begin{equation}\label{eq:WJSparameterized}
\begin{aligned}
    &  \tilde{W} = -  \tilde{T} \operatorname{Re}  \sum_{\omega_n > 0} \left[\cosh2\theta_{n} \cosh2\zeta_{n} - 1\right]\;,\\
    & \tilde{J}^S = - 2   \theta_{I} \tilde{T}   \operatorname{Im} \sum_{\omega_n > 0}  \sinh \zeta_{n}  \cosh ^2\zeta_{n}  \sinh \theta_{n}\;.
\end{aligned}
\end{equation}

\subsection{Solution procedure}
\label{subsec:Numerics_details}

In order to solve Eqs.~(\ref{eq:Algebraicparameterized}) and (\ref{eq:Selfconsistencydimensionless}),
we first fix $\tilde{p}^{S0}$ and $\tilde{\Delta}$. For fixed $\tilde{p}^{S0}$ and $\tilde{\Delta}$, we solve the algebraic Usadel equations, Eq.~(\ref{eq:Algebraicparameterized}), for every $\tilde{\omega}_n$ and obtain their solutions $\theta_n$, $\zeta_n$. With the help of these solutions, we update $\tilde{\Delta}$ for the given $\tilde{p}^{S0}$ till the selfconsistency in $\tilde{\Delta}$ is reached. Using $\theta_n$, $\zeta_n$, and $\tilde{\Delta}$ found in the previous stage, we compute $\tilde{J}^S(\tilde{p}^{S0})$ and $\tilde{W}(\tilde{p}^{S0})$ according to (\ref{eq:WJSparameterized}). Then $\tilde{I}_0(\tilde{p}^{S0}) = \tilde{J}^S(\tilde{p}^{S0}) + \tilde{W}(\tilde{p}^{S0}) \tilde{d}_S \tilde{p}^{S0}$ is used to calculate the diode efficiency, $\eta = \frac{I_{c+}-|I_{c-}|}{I_{c+}+|I_{c-}|}$, where $I_{c+} = \operatorname{max} \tilde{I}_0(\tilde{p}^{S0})$ and $I_{c-} = \operatorname{min} \tilde{I}_0(\tilde{p}^{S0})$.  By using this code for multiple values of the input parameters $\theta_{I}$, $\tilde{H}_{0}$ and $\tilde{\Gamma}^{\parallel}$, we generated Fig.~\ref{fig:diode_eta}.

During our numerical procedure, we use root finding routines provided by the Python library scipy.optimize \cite{virtanen2020scipy} to solve both the algebraic Usadel equation for each $\tilde{\omega}_n$ and the selfconsistency equation. In order to achieve a good speed of convergence for the sums over the Matsubara frequencies $\tilde{\omega}_n$, we employ the Gaussian sum quadrature proposed in \cite{virtanen2024numerical,monien2010gaussian},
\begin{equation}
    \sum_{\tilde{\omega}_n > 0} f(\tilde{\omega}_n) \approx \sum_{j=0}^N a_j f(\tilde{\omega}^\star_j)\;. 
\end{equation}
Here, the coefficients $a_j$ and the energies $\tilde{\omega}^\star_j$ are chosen in such a way that the equality is exact for any functions $f(\tilde{\omega}_n)$ that are polynomials in $(1 + n)^{-2}$ of order less than $N/2$. Here, $N$ is chosen so that $\tilde{\omega}_N$ is much larger than any energy scale in the system.

\end{document}